\documentclass[english,twocolumn,notitlepage]{revtex4-1}
\usepackage[T1]{fontenc}  \usepackage[latin1]{inputenc}
 \usepackage{babel} 
 \usepackage{epsfig,epsf,psfrag} 
\usepackage{graphicx} 
\usepackage{subfigure}
\usepackage{pslatex}
\usepackage{epic,eepic} \usepackage{bbold}
\usepackage{color,pstcol}\usepackage{amsmath}
\usepackage{pstricks} 
\usepackage{fancyhdr}  \parindent0em  

 \vspace{-10.0cm}  
\begin{document} \title{Disordered contacts can localize helical edge electrons} 
  \author{Arjun Mani}
 \author{Colin Benjamin} \email{colin.nano@gmail.com}\affiliation{School of Physical Sciences, National Institute of Science Education \& Research, HBNI, Jatni-752050, India}
\begin{abstract}
It is well known that quantum spin Hall (QSH) edge modes being helical are immune to backscattering due to non-magnetic disorder within the sample. Thus, quantum spin Hall edge modes are non-localized and show a vanishing Hall resistance along with quantized 2-terminal, longitudinal and non-local resistances even in presence of sample disorder. However, this is not the case for contact disorder. This paper shows that when all contacts are disordered in a N-terminal quantum spin Hall sample, then transport via these helical QSH edge modes can have a significant localization correction. All the resistances in a N-terminal quantum spin Hall sample deviate from their values derived while neglecting the phase acquired at disordered contacts, and this deviation is called the quantum localization correction. This correction term increases with the increase of disorderedness of contacts but decreases with the increase in number of contacts in a N terminal sample. The presence of inelastic scattering, however,  can completely destroy the quantum localization correction.  
\end{abstract}
 \maketitle
\section{Introduction}
The quantum spin Hall effect observed in a 2D topological insulator is known for transport via dissipation-less helical 1D edge modes. These 1D helical edge modes are robust to sample disorder and are observed in systems like HgTe/CdTe heterostructures at low temperatures, due to bulk spin orbit effects and in absence of a magnetic field \cite{asboth, sczhang, hasan}. QSH edge modes are helical, i.e., at the upper edge a spin-up electron moves in one direction while spin-down electron moves in opposite direction while at the lower edge the directions are reversed, see Fig.~1. Thus, quantum spin Hall systems are invariant under time reversal symmetry. Due to the topological nature of these edge modes, the Hall resistance vanishes, while the 2-terminal, longitudinal and non-local resistances are quantized at $\frac{3}{2}\frac{h}{2e^2}$, $\frac{1}{2}\frac{h}{2e^2}$ and $\frac{1}{6}\frac{h}{2e^2}$ respectively in a six terminal ideal QSH sample (without any disordered contacts). The Hall, longitudinal, 2-terminal and non-local conductances/resistances are determined by resorting to the Landauer-Buttiker(L-B) theory\cite{datta, sanvito}. In this formalism, for a QSH device with $N$ contacts, the current at contact $i$ at zero temperature is\cite{datta,sanvito,chulkov}:
\begin{eqnarray}
I_i&=&\frac{e^2}{h}\sum_{\substack{j=1 \\ j\neq i}}^N\sum_{\sigma,\sigma'}[T^{\sigma\sigma'}_{ji}V_i-T^{\sigma\sigma'}_{ij}V_j], \text{with }T^{\sigma\sigma'}_{ij}=Tr[s^{\sigma\sigma'\dagger}_{ij} s^{\sigma\sigma'}_{ij}],\nonumber\\
\end{eqnarray}

 \begin{figure}[h]
   \centering {\includegraphics[width=0.3\textwidth]{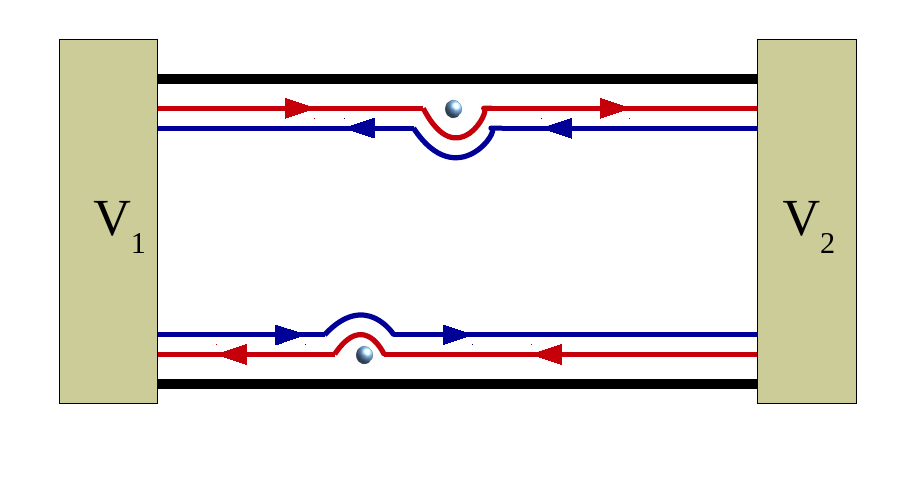}}\\
 \vspace{-.3cm}
\caption{Helical QSH edge modes are immune to sample disorder.}
\end{figure}

 \begin{figure}[h]
 \centering \subfigure[]{\includegraphics[width=0.43\textwidth]{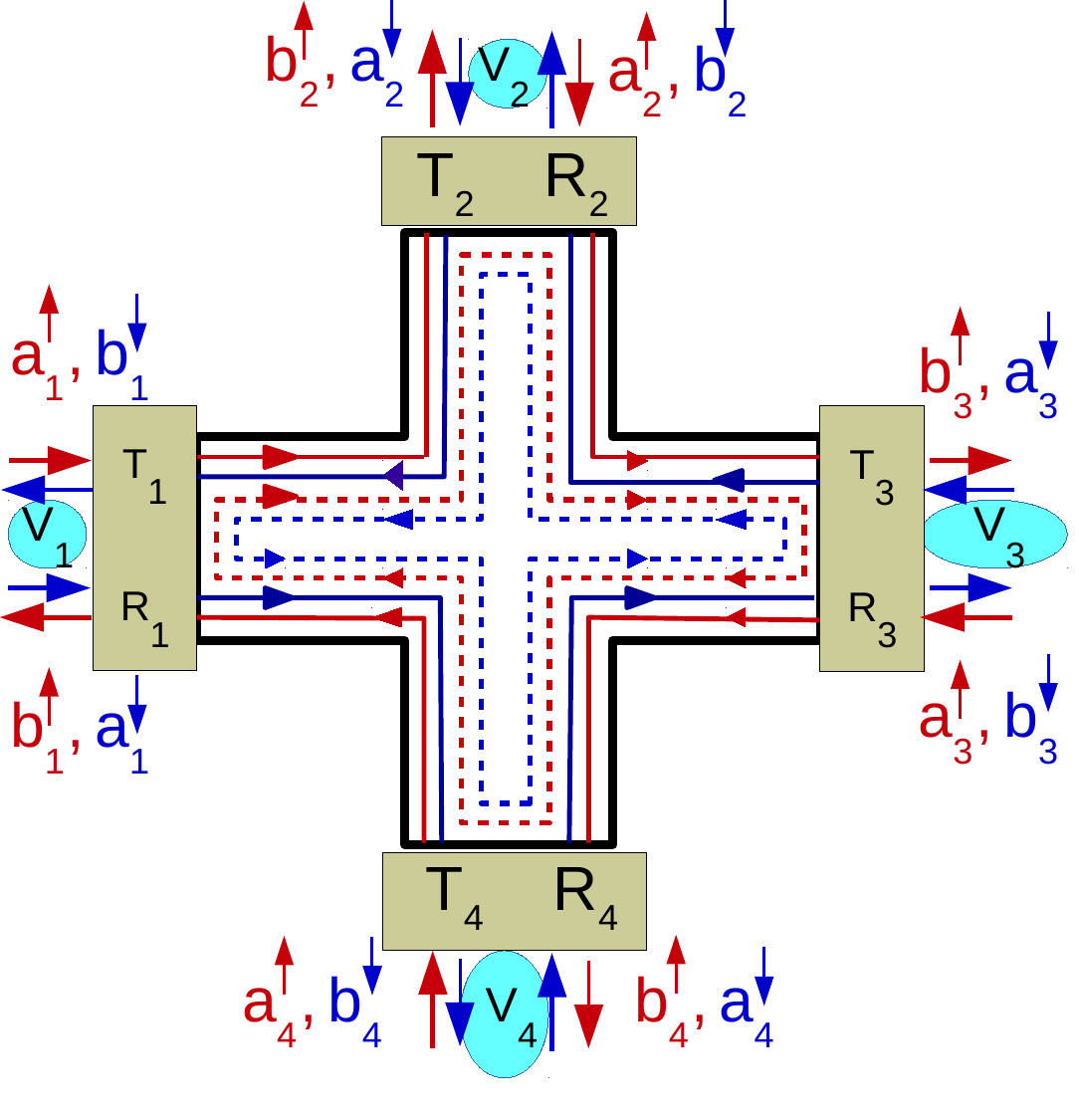}}\\
  \centering \subfigure[]{\includegraphics[width=0.43\textwidth]{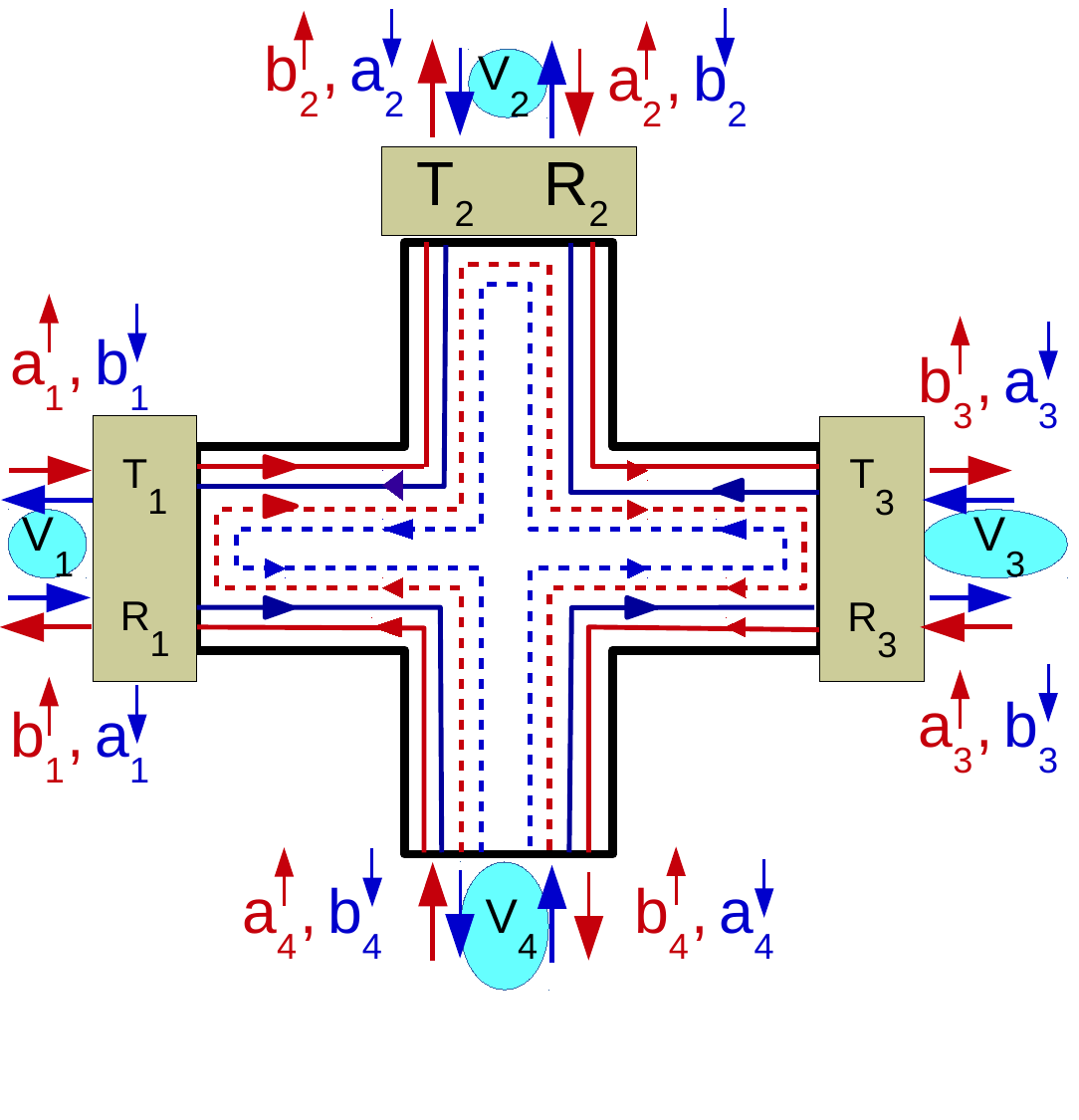}}\\
 \vspace{-.3cm}
\caption{(a) Pictorial representation to explain the origin of quantum localization correction when all contacts are disordered in a $4$T QSH sample. (b) Absence of quantum localization correction in a $4$T QSH sample when only $3$ of the $4$ contacts are disordered.}
\end{figure}
where $T^{\sigma\sigma'}_{ij}$ is the transmission probability for an electronic edge mode from contact $j$ to contact $i$ with initial spin $\sigma'$ to final spin $\sigma$, $V_i$ being the voltage bias applied at contact $i$, while $s^{\sigma\sigma'}_{ij}$ are the elements of the scattering matrix $\mathcal{S}$ of the $N$-terminal sample. 

\section{Motivation}
In quantum diffusive transport regime, localization of electronic states is well known \cite{al, datta}, the resistance of a sample increases exponentially with sample length ($l$) for $l>\xi$ ($\xi$ being localization length)~\cite{datta}. This is known as strong or Anderson localization\cite{jian, jain}. On the other hand, when the sample length $l\leq\xi$, the system shows an unique property: the resistance increases from the Ohmic result by universal factor $h/2e^2$. This increase by the universal factor $h/2e^2$ is called as weak localization correction. The QSH edge modes, as shown in Fig.~1, are immune to backscattering, e.g., if there is disorder in the sample (see, Fig.~1), edge modes will move around the disorder without their transmission probabilities getting affected due to topological protection. 
 \begin{figure*}
 \centering \subfigure[]{\includegraphics[width=0.35\textwidth]{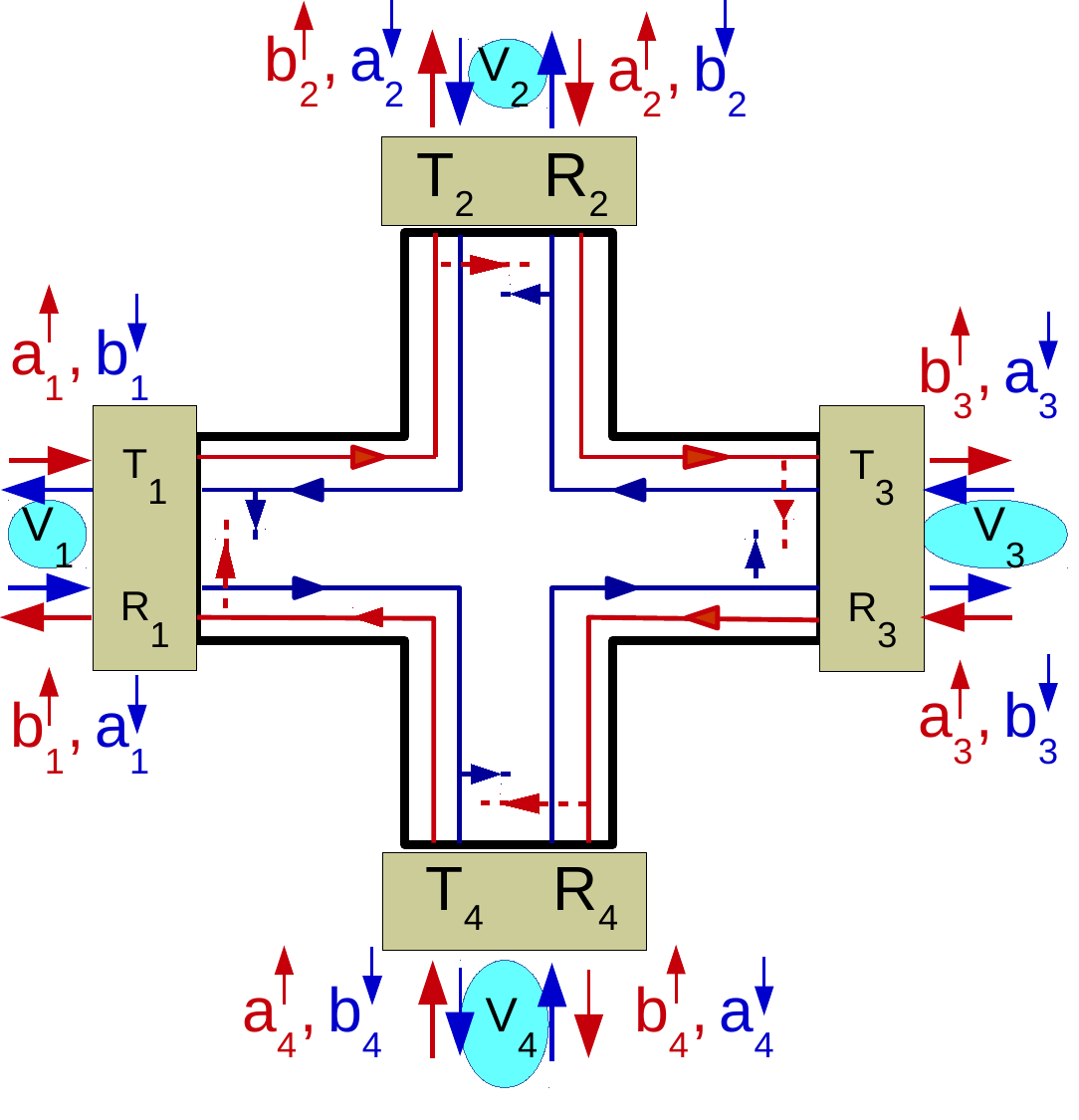}}
  \centering \subfigure[]{\includegraphics[width=0.45\textwidth]{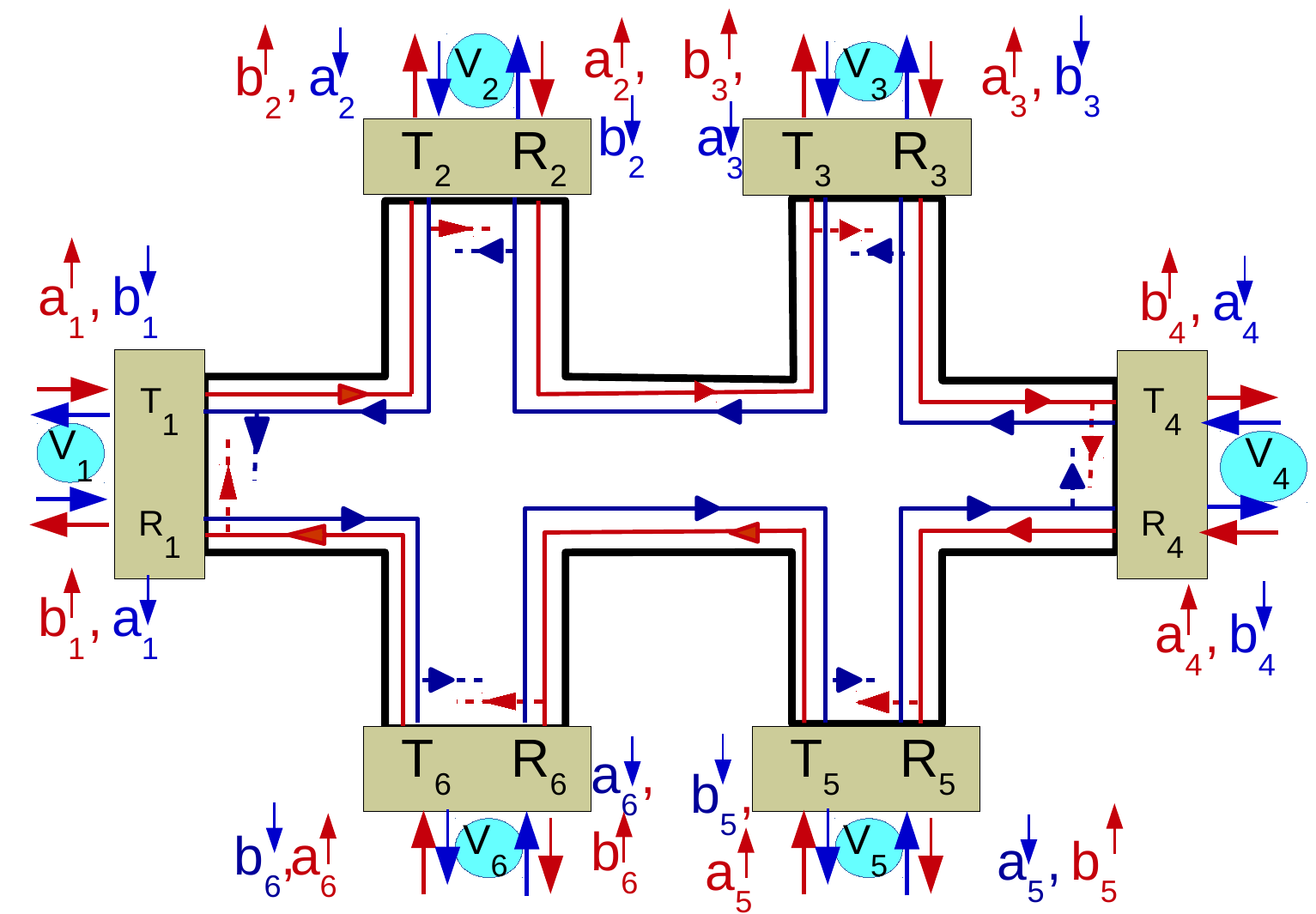}}
  \centering \subfigure[]{\includegraphics[width=.55\textwidth]{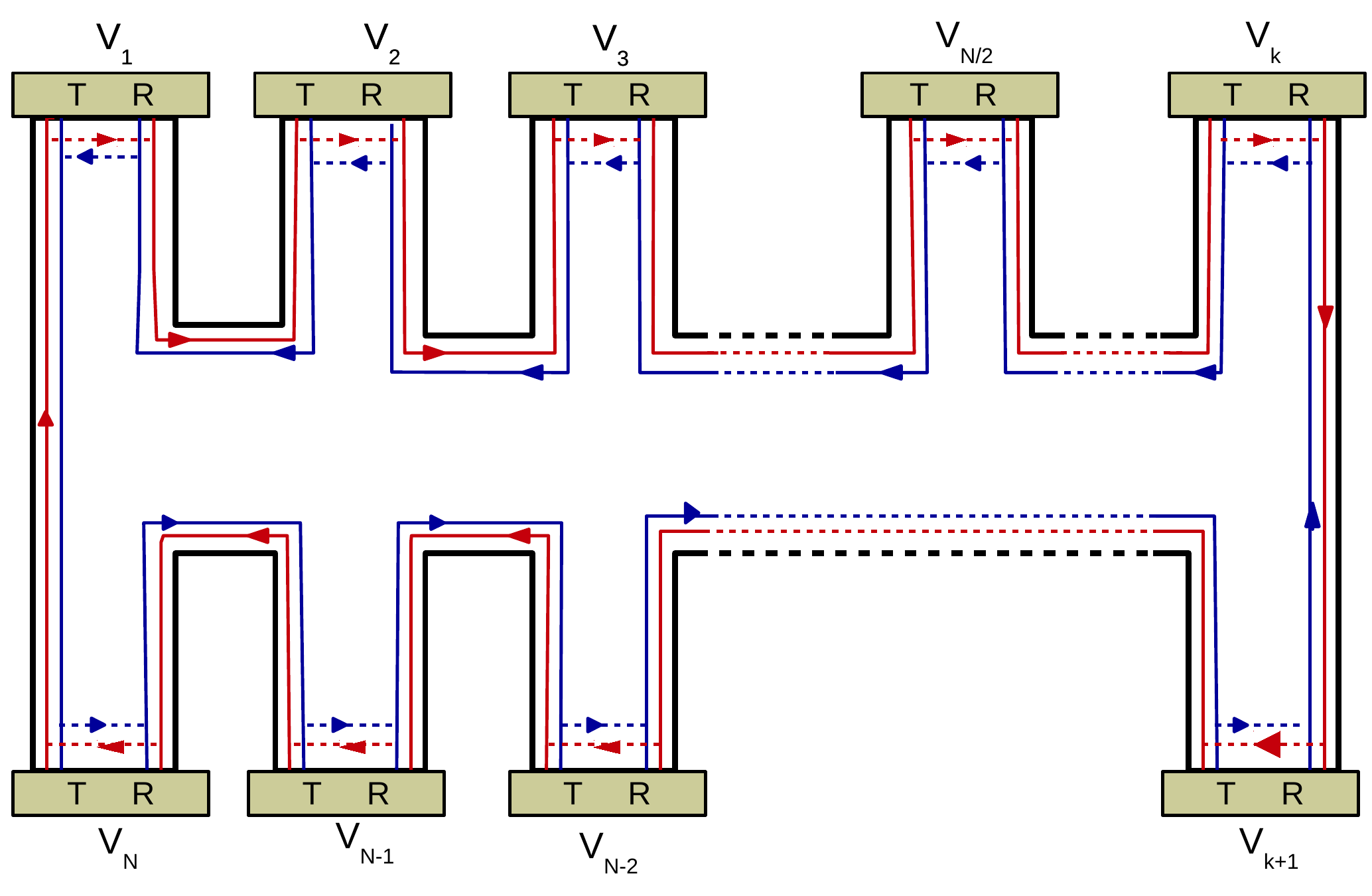}}
 \vspace{-.3cm}
\caption{(a) 4-terminal, (b) 6-terminal and (c) N-terminal QSH sample with all disordered contacts.}
\end{figure*}
In this work we however predict that, if a contact is disordered, i.e., can reflect edge modes partially then a ``quantum'' localization correction can arise for edge modes too but only when all contacts are disordered. What happens is backscattering of the electrons within the sample takes place when all contacts are disordered and thus multiple paths are generated from one contact to another. As a result, the transmission probabilities and resistances become dependent on the disorderedness of contacts. However, it should noted that this quantum localization observed for QSH edge modes is different from the weak localization correction seen in context of quantum diffusive transport. In quantum diffusive transport regime, the weak localization correction is universal ($h/2e^2$), while in our case, the correction due to localization as will be discussed in more detail in sections III and IV, depends on the strength of disorder at contacts and on the number of contacts. Further, this quantum localization correction is present only when all contacts are disordered, see Fig.~2(a). In Fig.~2(a), $a_i^\sigma$ and $b_i^\sigma$ refer to the incoming and outgoing edge states respectively from sample to contact $i$ with $\sigma$ being the spin index for that edge state. In Fig.~2(a), we see that a spin up electron in the $a_1^\uparrow$ edge state at contact $1$ can either transmit into the sample with probability $T_1$ or reflect back again to contact $1$ with probability $R_1$. After entering the sample, this edge state electron can reach contact $3$ via reflection at contact $2$ with probability $R_2$ and then transmit to contact $3$ with probability $T_3$. Thus the transmission probability for a spin-up edge electron from contact $1$ to $3$ is $T_1R_2T_3$. This is one among the infinite number of paths possible. For example, it can also reach contact $3$ by taking second path after reflecting at contacts $3, 4, 1, 2$ and then finally transmitting into contact $3$ with transmission probability $T_1T_3R_1R_2^2R_3R_4$. Thus, summing all paths from contact $1$ to $3$, we get the net transmission probability for the spin up edge state- $T^{\uparrow\uparrow}_{31}=T_1R_2T_3/(1-R_1R_2R_3R_4)$. However, by taking recourse to scattering amplitudes instead of probabilities we get the transmission amplitude from contact $1$ to $3$ as $t_{31}^{\uparrow\uparrow}=-t_1r_2t_3e^{i(\phi-\phi_4)}/(1-r_1r_2r_3r_4e^{i\phi})$, where $t_i$ and $r_i$ are the transmission and reflection amplitudes at contact $i$ with $\phi_i$ being the phase acquired by the electron at contact $i$ and $\phi=\sum_i\phi_i$. This scattering amplitude will lead to the transmission probability from contact $1$ to $3$ for spin up edge state- $T_{31}^{\uparrow\uparrow}=|t_{31}^{\uparrow\uparrow}|^2=T_1R_2T_3/(1+R_1R_2R_3R_4-2\sqrt{R_1R_2R_3R_4}\cos\phi)$. This is different to what was derived earlier for $T_{31}^{\uparrow\uparrow}$. Similarly, rest of the transmission probabilities can be calculated by considering transmission probabilities or via following scattering amplitudes, and these too will be different for each case. Thus, when an infinite number of paths exist from one contact to another then a difference between the average resistances derived from scattering amplitudes $\langle R^{Amp}_X\rangle$ (wherein $X=H,L,2T,NL$ denotes Hall, Longitudinal, Two-terminal and Non-local) and resistances derived from probabilities $R^{}_X$, i.e., $\langle R^{Amp}_X\rangle\neq R^{}_X$ is seen. This situation changes, if however at least one of the contacts is not disordered, see Fig.~2(b) (wherein contact $4$ is not disordered), in this case there are a finite number of paths from one contact to another. This can be seen as follows: in Fig.~2(b), a spin up edge state from contact $1$ can reach contact $3$ by following only one path via reflection at contact $2$ with probability $T_{31}^{\uparrow\uparrow}= T_1R_2T_3$. There is no second path to reach contact $3$, since contact $4$ is not disordered, this edge state can not reflect from contact $4$. Further, the scattering amplitude 
from contact $1$ to $3$ is $t_{31}^{\uparrow\uparrow}=-t_1r_2t_3e^{i(\phi-\phi_4)}$, 
which gives the transmission probability $T_{31}^{\uparrow\uparrow}=|t_{31}^{\uparrow\uparrow}|^2=T_1R_2T_3$. Thus the calculation using scattering probabilities and that with scattering amplitudes yield identical results for the case when less than $N$ contacts are disordered. This results in $\langle R^{Amp}_X\rangle= R^{}_X$ for the case when less than $N$ contacts are disordered and thus quantum localization correction vanishes. Similar, to what is described here for QSH system, was also shown recently for quantum Hall (QH) system in Ref.~\cite{arjun3}. This is the main motivation of our work, can we see a similar quantum localization correction for QSH samples? Since QSH edge modes are helical (spin polarized) rather than chiral (spin unpolarized) as in QH sample, it will be interesting to see the effect of spin polarized and helical edge modes on the quantum localization correction. Further, to compare the characteristics of this quantum localization correction in various resistances for both QH and QSH systems is another motivation of this paper. We elaborate on this in sections III, IV and V for four, six and N-terminal QSH samples respectively.
 The topic of  research undertaken in this paper is both timely as well as novel. Since understanding why in quantum spin Hall experiments the robust quantized conductance is absent is a hotly debated topic of research. Reasons for the less than robust quantization of spin Hall conduction have ranged from magnetic impurities to inelastic scattering as well as to hyperfine interaction which will break time reversal symmetry and therefore induce backscattering of edge modes\cite{vayrynen}. {In this manuscript, we show that} even when either there is no inelastic scattering  or in absence of magnetic impurities or even for no hyperfine interaction\cite{hsu} there still can be loss in quantized conductance which we call a quantum localization correction due to disordered contacts alone. Further all the proposals to explain the loss of quantization of helical conduction in quantum spin Hall samples rely on some kind of inelastic scattering which is dealt with via many body interactions. Our paper is novel in that we via a single particle theory explain the loss of quantized conduction which we dub the quantum localization correction to helical edge transport. 

The organization of this paper is as follows: in section III, we deal with a 4-terminal QSH sample with all disordered contacts and derive an expression for the quantum localization correction, while in sections IV and V we discuss the six and N-terminal QSH samples. Next in section VI, we study the impact of inelastic scattering on this quantum localization correction. We conclude with a table summarizing the main results of our paper and compare it with results derived in Ref.~\cite{arjun3}. 
 
 \section{Four terminal system with all disordered contacts }
A 4-terminal QSH sample is shown in Fig.~3(a) with all disordered contacts. The strength of disorder at contact $i$ is defined by $D_i$ and it is related to the reflection ($R_i$) and transmission probabilities ($T_i$) of an edge state at contact $i$ by the relation $D_i=R_i=1-T_i$. Contacts $1$, $3$ are current probes while contacts $2$, $4$ are voltage probes, such that $I_2=I_4=0$. For calculating the current at each of these contacts, we need to derive the edge state transmission probability $T^{\sigma\sigma'}_{ij}$ between these contacts. Since all contacts are disordered, we need to consider the scattering amplitudes to calculate the transmission probabilities $T^{\sigma\sigma'}_{ij}$, from Eq.~(1). First we write down the scattering matrix $S_j$ at each contact $j$ separately relating incoming edge modes ($a^\uparrow_j,a^\downarrow_j, a'^{\uparrow}_j,a'^{\downarrow}_j$) to outgoing edge modes ($b^\uparrow_j,b^\downarrow_j, b'^{\uparrow}_j,b'^{\downarrow}_j$) at that particular contact $j$ and then deduce the full scattering matrix $\mathcal{S}$ of the system out of the contact scattering matrices $S_j$, see Ref.~\cite{arjun4}. The scattering matrix $S_j$ is defined as follows
 \begin{eqnarray}
 S_j=\begin{pmatrix}
 r_je^{i\phi_j^{r,\uparrow}}&0&t_je^{i\phi_j^{t,\uparrow}}&0\\
  0&r_je^{i\phi_j^{r,\downarrow}}&0&t_je^{i\phi_j^{t,\downarrow}}\\
  t_je^{i\phi_j^{t,\uparrow}}&0&r_je^{i\phi_j^{r,\uparrow}}&0\\
0& t_je^{i\phi_j^{t,\downarrow}}&0&r_je^{i\phi_j^{r,\downarrow}}
 \end{pmatrix},
 \end{eqnarray}
 where $r_j$ and $t_j$ are the reflection and transmission amplitudes respectively at contact $j$, $\phi^{r,\sigma}_j$ and $\phi^{t,\sigma}_j$ are the reflection and transmission phase acquired by the spin $\sigma$(=$\uparrow/\downarrow$) edge electron via scattering at the disordered contact $j$. Unitarity of the scattering matrix $S_j$ dictates $S_j^\dagger S_j=S_jS_j^\dagger=\mathbb{I}$, which implies- $\phi_j^{r,\sigma}=\phi_j^{t,\sigma}-\frac{\pi}{2}=\phi_j$, (dropping the spin index $\sigma$ from the phase as disorder is spin independent). Thus the scattering matrix $S_j$ reduces to
  \begin{eqnarray}
 S_j=\begin{pmatrix}
 r_je^{i\phi_j}&0&it_je^{i\phi_j}&0\\
  0&r_je^{i\phi_j}&0&it_je^{i\phi_j}\\
  it_je^{i\phi_j}&0&r_je^{i\phi_j}&0\\
0& it_je^{i\phi_j}&0&r_je^{i\phi_j}
 \end{pmatrix}.
 \end{eqnarray}
Each element of the full scattering matrix $\mathcal{S}$ can be calculated from these $S_j$ matrices in the following manner: an electron in $a_1$ edge state can scatter into $b_1$ edge state directly with amplitude $r_1e^{i\phi_1}$, but then, it can also follow a different path via scattering at contacts $2,3,4$ and reach $b_1$ edge state with amplitude: $it_1e^{i\phi_1}\times r_2e^{i\phi_2}\times r_3e^{i\phi_3}\times r_4e^{i\phi_4}\times it_1e^{i\phi_1} =-t_1^2r_2r_3r_4e^{i(2\phi_1+\phi_2+\phi_3+\phi_4)}$, and a third path with amplitude: $-t_1^2r_1(r_2r_3r_4)^2e^{i(3\phi_1+2\phi_2+2\phi_3+2\phi_4)}$ and so on. Summing over all these paths we get $(1,1)^{\mbox{th}}$ element $s_{11}^{\uparrow\uparrow}$ of full scattering matrix $\mathcal{S}$ of the system, which is $(r_1-r_2r_3r_4e^{i\phi})e^{i\phi_1}/(1-r_1r_2r_3r_4e^{i\phi})$, with $\phi=\phi_1+\phi_2+\phi_3+\phi_4$. Similarly, rest of the elements of the $\mathcal{S}$ matrix can be derived. The scattering matrix for 4-terminal QSH sample in Fig.~3(a) is thus:
\begin{widetext}
\begin{equation} 
 \mathcal{S}=\frac{1}{a}\left(\begin{smallmatrix}
 (r_1-r_2r_3r_4e^{i\phi})e^{i\phi_1}&0&-t_1t_2r_3r_4e^{i\phi}&0&-t_1t_3r_4e^{i(\phi-\phi_2)}&0&-t_1t_4e^{i(\phi_1+\phi_4)}&0\\
 0&(r_1-r_2r_3r_4e^{i\phi})e^{i\phi_1}&0&-t_1t_2e^{i(\phi_1+\phi_2)}&0&-t_1t_3r_2e^{i(\phi-\phi_4)}&0&-t_1t_4r_2r_3e^{i\phi}\\
 -t_1t_2e^{i(\phi_1+\phi_2)}&0&(r_2-r_1r_3r_4e^{i\phi})e^{i\phi_2}&0&-t_2t_3r_1r_4e^{i\phi}&0&-t_2t_4r_1e^{i(\phi-\phi_3)}&0\\
0& -t_1t_2r_3r_4e^{i\phi}&0&(r_2-r_1r_3r_4e^{i\phi})e^{i\phi_2}&0&-t_2t_3e^{i(\phi_2+\phi_3)}&0&-t_2t_4r_3e^{i(\phi-\phi_1)}\\
 -t_1t_3r_2e^{i(\phi-\phi_4)}&0&-t_2t_3e^{i(\phi_2+\phi_3)}&0&(r_3-r_1r_2r_4e^{i\phi})e^{i\phi_3}&0&-t_3t_4r_1r_2e^{i\phi}&0\\
 0&-t_1t_3r_4e^{i(\phi-\phi_2)}&0&-t_2t_3r_1r_4e^{i\phi}&0&(r_3-r_1r_2r_4e^{i\phi})e^{i\phi_3}&0&-t_3t_4e^{i(\phi_3+\phi_4)}\\
 -t_1t_4r_3r_2e^{i\phi}&0&-t_2t_4r_3e^{i(\phi-\phi_1)}&0&-t_3t_4e^{i(\phi_3+\phi_4)}&0&(r_4-r_1r_2r_3e^{i\phi})e^{i\phi_4}&0\\
 0& -t_1t_4e^{i(\phi_1+\phi_4)}&0&-t_2t_4r_1e^{i(\phi-\phi_3)}&0&-t_3t_4r_1r_2e^{i\phi}&0&(r_4-r_1r_2r_3e^{i\phi})e^{i\phi_4}\end{smallmatrix}\right),
 \end{equation}
 \end{widetext}
 where $a=1-r_1r_2r_3r_4e^{i\phi}$. This full scattering matrix $\mathcal{S}$ relates the incoming edge modes to the outgoing edge modes (see, Fig.~3(a)) of the system via the relation $(b_1^\uparrow,b_1^\downarrow,b_2^{\uparrow},b_2^\downarrow,b_3^\uparrow,b_3^\downarrow,b_4^\uparrow,b_4^\downarrow)^T=\mathcal{S}(a_1^\uparrow,a_1^\downarrow,a_2^{\uparrow},a_2^\downarrow,a_3^\uparrow,a_3^\downarrow,a_4^\uparrow,a_4^\downarrow)^T$. Current conservation is guaranteed by the unitarity of the scattering matrix $\mathcal{S}$. The conductance matrix $G$ of the system deduced from the full scattering matrix $\mathcal{S}$, following from Eq.~(1), is
 {\small
\begin{eqnarray} 
 G=\frac{e^2}{h}\frac{1}{a'}\left(\begin{smallmatrix}
2(1-R_2R_3R_4)T_1&-T_1T_2(R_3R_4+1)&-T_1T_3(R_2+R_4)&-T_1T_4(1+R_2R_3)\\
 -T_1T_2(1+R_3R_4)&2(1-R_1R_3R_4)T_2&-T_2T_3(R_1R_4+1)&-T_2T_4(R_1+R_3)\\
 -T_1T_3(R_2+R_4)&-T_2T_3(1+R_1R_4)&2(1-R_1R_2R_4)T_3&-T_3T_4(R_1R_2+1)\\
 -T_1T_4(1+R_3R_2)&-T_2T_4(R_1+R_3)&-T_3T_4(1+R_1R_2)&2(1-R_1R_2R_3)T_4\end{smallmatrix}\right),\nonumber\\
 \end{eqnarray}
}
where $a'=(1+R_1R_2R_3R_4-2\sqrt{R_1R_2R_3R_4}\cos\phi)$. Conductance matrix $G$ connects currents and voltages at each contact via the relation $(I_1,I_2,I_3,I_4)^T=G(V_1,V_2,V_3,V_4)^T$. Since currents through voltage probes $2$ and $4$ are zero, so $I_2=I_4=0$, and choosing reference potential $V_3=0$ we get voltages $V_2$ and $V_4$ in terms of $V_1$. Hall resistance $R^{Amp}_H=R_{13,24}=\frac{(V_2-V_4)}{I_1}$, 2-terminal resistance is $R^{Amp}_{2T}=R_{13,13}=\frac{(V_1-V_3)}{I_1}$, and non-local resistance is $R^{Amp}_{NL}=R_{12,43}=\frac{(V_4-V_3)}{I_1}$ (to calculate the non-local resistance contacts $1,2$ are used as current probes while contacts $3,4$ as voltage probes). Here, we consider $D_1=D_2=D_u$, and $D_3=D_4=D_l$ for Hall resistance since for equally disordered contacts it vanishes. For 2-terminal and non-local, we consider $D_u=D_l=D$. Thus,   
\begin{eqnarray}
R^{Amp}_H&=&\frac{h}{2e^2}\frac{(D_l - D_u)^2 (1 + D_l^2 D_u^2 - 2 D_l D_u \cos\phi)}{(4 (1 + 
   D_l) (1 + D_u) (-1 + D_l D_u)^2 (1 + D_l D_u))},\nonumber\\
R^{Amp}_{2T}&=&\frac{h}{2e^2}\frac{(1 + D^4 - 2 D^2 \cos\phi)}{(1 - D^2)^2},\nonumber\\
\text{and }R^{Amp}_{NL}&=&\frac{h}{2e^2}\frac{(1 + D^4 - 2 D^2 \cos\phi)}{(1 + D^2)(1+D)^2}.
\end{eqnarray}
The mean Hall, 2-terminal and non-local resistances obtained by averaging over the phase $'\phi'$ acquired by the electronic edge modes due to multiple scattering at disordered contacts is thus- 
\begin{eqnarray}
\langle R^{Amp}_H\rangle&=&\frac{h}{2e^2}\frac{((D_l - D_u)^2 (1 + D_l^2 D_u^2))}{(4 (1 + 
   D_l) (1 + D_u) (-1 + D_l D_u)^2 (1 + D_l D_u))},\nonumber\\
\langle R^{Amp}_{2T}\rangle&=&\frac{h}{2e^2}\frac{(1 + D^4)}{(1 - D^2)^2},\nonumber\\
\langle R^{Amp}_{NL}\rangle&=&\frac{h}{2e^2}\frac{(1 + D^4)}{(1 + D^2)(1+D)^2}.
\end{eqnarray}
One observes that the mean Hall, 2-terminal and non-local resistances lose their quantization due to disordered contacts. To calculate the quantum localization correction, we need to calculate the resistances using probabilities ignoring the phases acquired by edge modes at disordered contacts. The conductance matrix $G$ is then

{\scriptsize
\begin{equation} 
 G=\frac{2e^2}{h}\frac{1}{a''}\begin{pmatrix}
(1-R_2R_3R_4)T_1&-T_1T_2R_3R_4&-T_1T_3T_4&-T_1T_4\\
 -T_1T_2&(1-R_1R_3R_4)T_2&-T_2T_3R_1R_4&-T_2T_4R_1\\
 -T_1T_3R_2&-T_2T_3&(1-R_1R_2R_4)T_3&-T_3T_4R_1R_2\\
 -T_1T_4R_3R_2&-T_2T_4R_3&-T_3T_4&(1-R_1R_2R_3)T_4\end{pmatrix},
 \end{equation}}
where $a''=(1-R_1R_2R_3R_4)$. As before,  the current through voltage probes $2,4$ is zero, and the reference potential $V_3=0$. Thus, the potentials $V_2$ and $V_4$ are derived in terms of $V_1$. The Hall resistance $R^{}_H$, 2-terminal resistance $R^{}_{2T}$, and nonlocal resistance $R^{}_{NL}$ calculated via probabilities are then-
 \begin{eqnarray}
R^{}_H&=&\frac{h}{e^2}\frac{(D_u - D_l)^2}{4 (1 + D_l) (1 + D_u) (1 - D_l D_u)},
\text{ } R^{}_{2T}=\frac{h}{e^2}\frac{1+D^2}{1-D^2},\nonumber\\ \text{and }R^{}_{NL}&=&\frac{h}{e^2}\frac{1-D}{1+D}.
\end{eqnarray}
The quantum localization correction is defined as the difference in the resistances calculated from amplitudes and that from probabilities, is then $R^{QL}_X=\langle R^{Amp}_X\rangle-R_X^{}$, with $X=H, 2T, NL$-
\begin{eqnarray}
R^{QL}_H&=&\frac{h}{2e^2}\frac{D_l^2D_u^2(D_l-D_u)^2}{2(1+D_l)(1+D_u)(1-D_l^2D_u^2)},\nonumber\\
R^{QL}_{2T}&=&\frac{h}{2e^2}\frac{2D^4}{(1-D^2)^2}, \quad
R^{QL}_{NL}=\frac{h}{2e^2}\frac{D^4}{2(1+D)^2(1+D^2)}.
\end{eqnarray}
It should be noted here, if $D_u=D_l$, then the quantum localization correction in case of Hall resistances $R^{QL}_H$ vanishes for equally disordered contacts. However, quantum localization correction does not vanish for 2-terminal and non-local resistances for equally disordered contacts. For 2-terminal and non-local resistances the quantum localization correction increases as disorder increases. Here, we also see from Eq.~(10) that if at least one of the contacts is not disordered, i.e., $D_i=0$ for either contacts $i=1$ or $2$ or $3$ or $4$ then quantum localization correction vanishes for 2-terminal and non-local resistances too (the factor $D^4$ in the numerator in the expression of 2-terminal and non-local resistances of Eq.~(10) comes from the product of $D_1$, $D_2$, $D_3$ and $D_4$, i.e., $D^4=D_1D_2D_3D_4$, when all contacts are equally disordered). Thus the 2-terminal and non-local resistances calculated via probabilities and via amplitudes are identical for the case when one contact is not disordered. This condition holds for any number of contacts as shown in the following sections.
 \section{Six terminal QSH system with all  disordered contacts}
Fig.~3(b) shows a 6-terminal QSH sample with all disordered contacts. Contacts $1, 4$ are used as current probes while $2, 3, 5, 6$ are used as voltage probes, such that current through these contacts is zero, i.e., $I_2=I_3=I_5=I_6=0$. The scattering matrix of the system shown in Fig.~3(b) is
\begin{widetext}
{\small
\begin{eqnarray} 
 \mathcal{S}=\frac{1}{b}\left(\begin{smallmatrix}
 (r-r^5e^{i\phi})e^{i\phi_1}&0&-t^2r^4e^{i\phi}&0&-t^2r^3e^{i\phi_{34561}}&0&-t^2r^2e^{i\phi_{4561}}&0&-t^2re^{i\phi_{561}}&0&-t^2e^{i\phi_{61}}&0\\
0&(r-r^5e^{i\phi})e^{i\phi_1}&0&-t^2e^{i\phi_{12}}&0&-t^2re^{i\phi_{123}}&0&
-t^2r^2e^{i\phi_{1234}}&0&-t^2r^3e^{i\phi_{12345}}&0&-t^2r^4e^{i\phi}\\
 -t^2r^2e^{i\phi_{12}}&0&(r-r^5e^{i\phi})e^{i\phi_2}&0&-t^2r^4e^{i\phi}&0&-t^2r^3e^{i\phi_{45612}}&0&-t^2r^2e^{i\phi_{5612}}&0&-t^2re^{i\phi_{612}}&0\\
 0& -t^2r^4e^{i\phi}&0&(r-r^5e^{i\phi})e^{i\phi_2}&0&-t^2e^{i\phi_{23}}&0&-t^2re^{i\phi_{234}}&0&-t^2r^2e^{i\phi_{2345}}&0&-t^2r^3e^{i\phi_{23456}}\\
 -t^2re^{i\phi_{123}}&0&-t^2e^{i\phi_{23}}&0&(r-r^5e^{i\phi})e^{i\phi_3}&0&-t^2r^4e^{i\phi}&0&-t^2r^3e^{i\phi_{56123}}&0&-t^2r^2e^{i\phi_{6123}}&0\\
 0& -t^2r^3e^{i\phi_{34561}}&0&-t^2r^4e^{i\phi}&0&(r-r^5e^{i\phi})e^{i\phi_3}&0&-t^2e^{i\phi_{34}}&0&-t^2re^{i\phi_{345}}&0&-t^2r^2e^{i\phi_{3456}}\\
 -t^2r^2e^{i\phi_{1234}}&0&-t^2re^{i\phi_{234}}&0&-t^2e^{i\phi_{34}}&0&(r-r^5e^{i\phi})e^{i\phi_4}&0&-t^2r^4e^{i\phi}&0&-t^2r^3e^{i\phi_{61234}}&0\\
  0&-t^2r^2e^{i\phi_{4561}}&0&-t^2r^3e^{i\phi_{45612}}&0&-t^2r^4e^{i\phi}&0&(r-r^5e^{i\phi})e^{i\phi_4}&0&-t^2e^{i\phi_{45}}&0&-t^2re^{i\phi_{456}}\\
-t^2r^3e^{i\phi_{12345}}&0&-t^2r^2e^{i\phi_{2345}}&0&-t^2re^{i\phi_{345}}&0&-t^2e^{i\phi_{45}}&0&(r-r^5e^{i\phi})e^{i\phi_5}&0&-t^2r^4e^{i\phi}&0\\
  0&-t^2e^{i\phi_{561}}&0&-t^2r^2e^{i\phi_{5612}}&0&-t^2r^3e^{i\phi_{56123}}&0&-t^2r^4e^{i\phi}&0&(r-r^5e^{i\phi})e^{i\phi_5}&0&-t^2e^{i\phi_{56}}\\
 -t^2r^4e^{i\phi}&0&-t^2r^3e^{i\phi_{23456}}&0&-t^2r^2e^{i\phi_{3456}}&0&-t^2re^{i\phi_{456}}&0&-t^2e^{i\phi_{56}}&0&(r-r^5e^{i\phi})e^{i\phi_6}&0\\
  0&-t^2e^{i\phi_{16}}&0&-t^2re^{i\phi_{612}}&0&-t^2r^2e^{i\phi_{6123}}&0&-t^2r^3e^{i\phi_{61234}}&0&-t^2r^4e^{i\phi}&0&(r-r^5e^{i\phi})e^{i\phi_6}\end{smallmatrix}\right),\nonumber\\
 \end{eqnarray}
}\end{widetext}
where $b=1-r^6e^{i\phi}$ with $\phi_{ij..m}=\phi_i+\phi_j+..+\phi_m$. For simplicity, we consider all contacts to be equally disordered. $r$ and $t$ denote reflection and transmission amplitudes at contact $i$. Scattering matrix $\mathcal{S}$ of the 6-terminal QSH sample relates the incoming spin-polarized edge states to the outgoing states (see, Fig.~3(b)) of the system via the relation: $(b_1^\uparrow,b_1^\downarrow,b_2^\uparrow,b_2^\downarrow,b_3^\uparrow,b_3^\downarrow,b_4^\uparrow,b_4^\downarrow,b_5^\uparrow,
b_5^\downarrow,b_6^\uparrow,b_6^\downarrow)^T=\mathcal{S} (a_1^\uparrow,a_1^\downarrow,a_2^\uparrow,a_2^\downarrow,a_3^\uparrow,a_3^\downarrow
,a_4^\uparrow,a_4^\downarrow,a_5^\uparrow,a_5^\downarrow,a_6^\uparrow,a_6^\downarrow)^T$. The conductance matrix $G$ of the sample deduced from scattering matrix $\mathcal{S}$ of Eq.~(11), and using Eq.~(1), is
{\small
\begin{eqnarray} 
 G=\frac{2e^2}{h}\frac{1}{b'}\left(\begin{smallmatrix}
2(1-R^5)T&-T^2R^4&-T^2R^3&-T^2R^2&-T^2R&-T^2\\
 -T^2&2(1-R^5)T&-T^2R^4&-T^2R^3&-T^2R^2&-T^2R\\
 -T^2R&-T^2&2(1-R^5)T&-T^2R^4&-T^2R^3&-T^2R^2\\
 -T^2R^2&-T^2R&-T^2&2(1-R^5)T&-T^2R^4&-T^2R^3\\
 --T^2R^3&-T^2R^2&-T^2R&-T^2&2(1-R^5)T&-T^2R^4\\
 -T^2R^4&-T^2R^3&-T^2R^2&-T^2R&-T^2&2(1-R^5)T\end{smallmatrix}\right),\nonumber\\
 \end{eqnarray}
}
where $b'=(1+R^6-2R^3\cos\phi)$. Since current through voltage probes $2,3,5$ and $6$ is zero, so $I_2=I_3=I_5=I_6=0$, and choosing reference potential $V_4=0$ we get potentials $V_2$, $V_3$, $V_5$ and $V_6$ in terms of $V_1$. Thus, the Hall resistance $R^{Amp}_H=R_{14,26}=\frac{(V_2-V_6)}{I_1}$, 2-terminal resistance $R^{Amp}_{2T}=R_{14,14}=\frac{(V_1-V_4)}{I_1}$, longitudinal resistance $R^{Amp}_L=R_{14,23}=\frac{(V_2-V_3)}{I_1}$, and non-local resistance $R^{Amp}_{NL}=R_{12,54}=\frac{(V_5-V_4)}{I_1}$ (to calculate non-local resistance, as before, contacts $1,2$ are used as current probes while contacts $3,4,5,6$ are voltage probes) are-
  \begin{figure*}
  \centering \subfigure[]{ \includegraphics[width=0.33\textwidth]{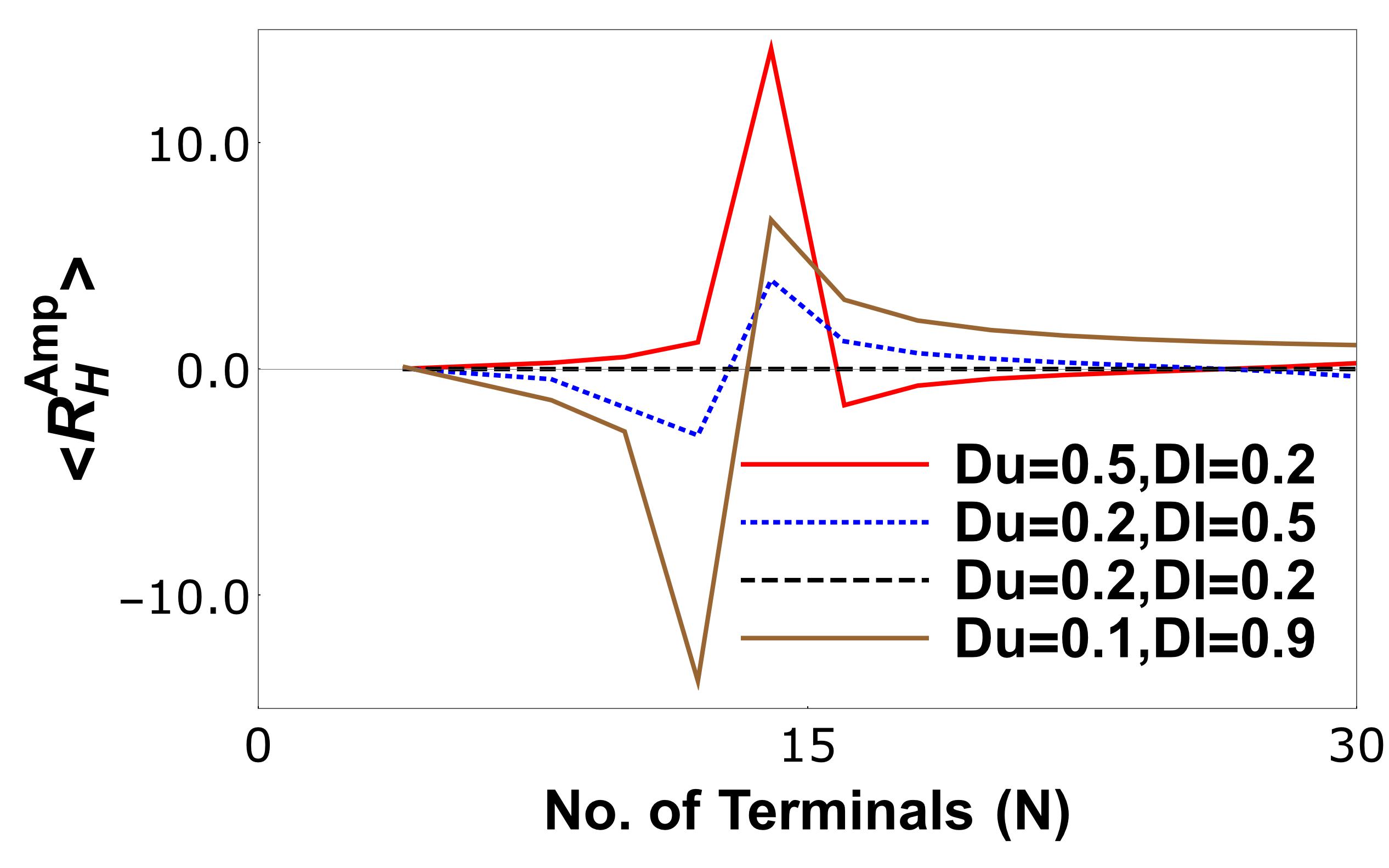}}
 \centering    \subfigure[]{ \includegraphics[width=0.32\textwidth]{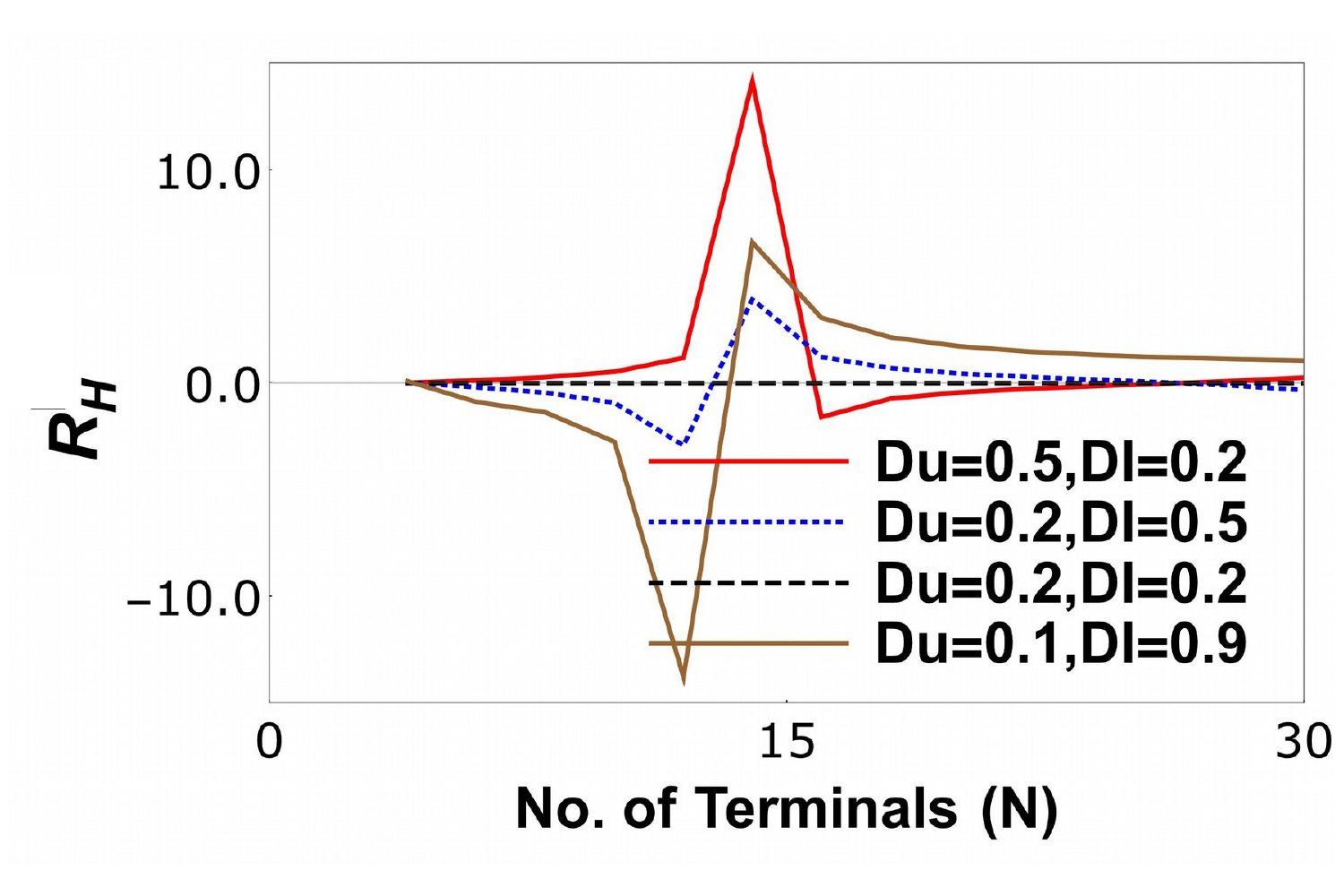}}
 \centering \subfigure[]{\includegraphics[width=.33 \textwidth]{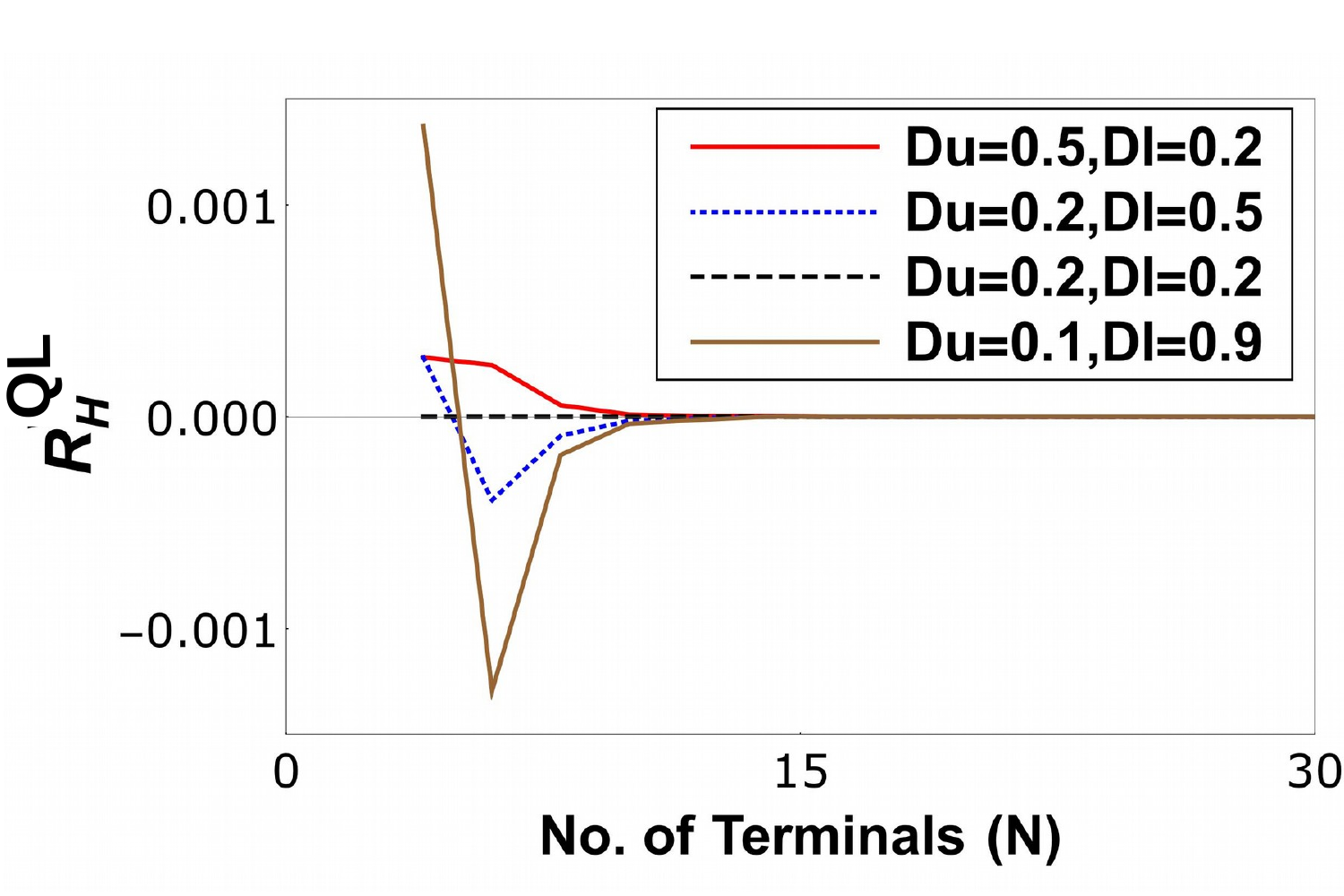}}
\vskip -0.2 in \caption{Hall resistance in units of $\frac{h}{e^2}$ calculated (a) via scattering amplitudes and (b) via probabilities for a N terminal QSH sample with lower edge contacts disordered with strength $D_l$ and upper edge contacts disordered with strength $D_u$, (c) the quantum localization correction to the Hall resistance.}
\end{figure*} 
 
 \begin{figure*}
  \centering \subfigure[]{ \includegraphics[width=0.33\textwidth]{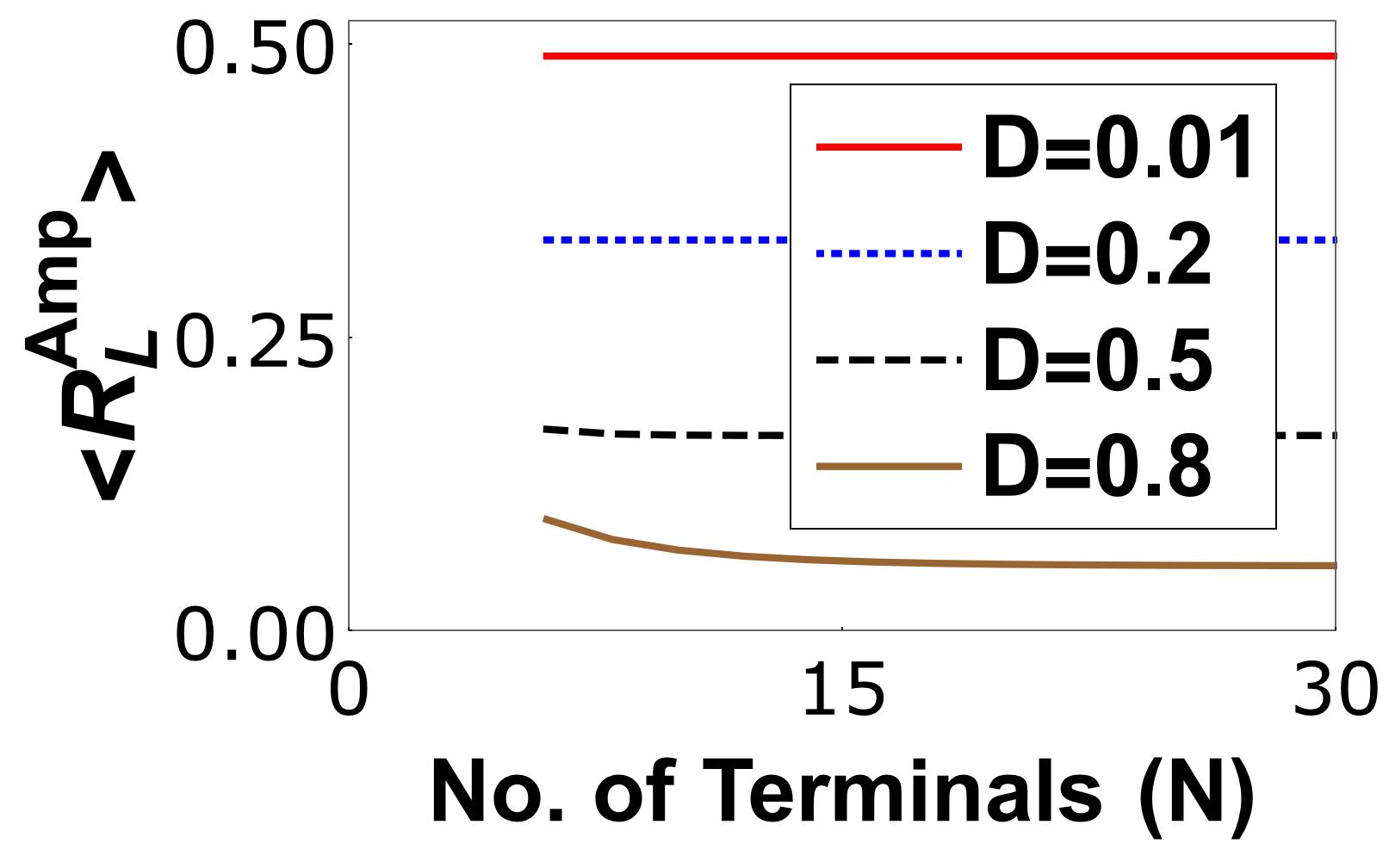}}
 \centering    \subfigure[]{ \includegraphics[width=0.32\textwidth]{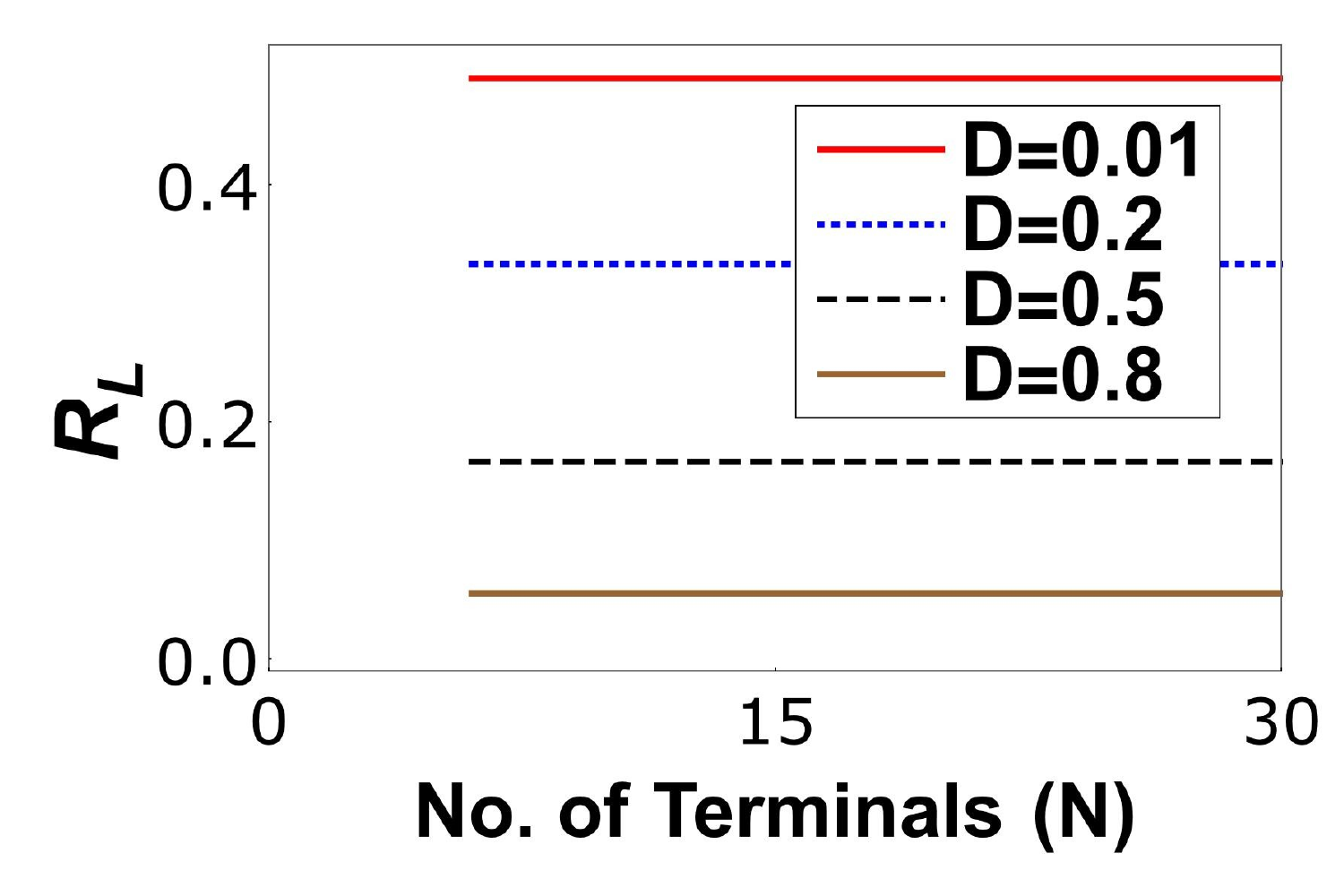}}
 \centering \subfigure[]{\includegraphics[width=.33 \textwidth]{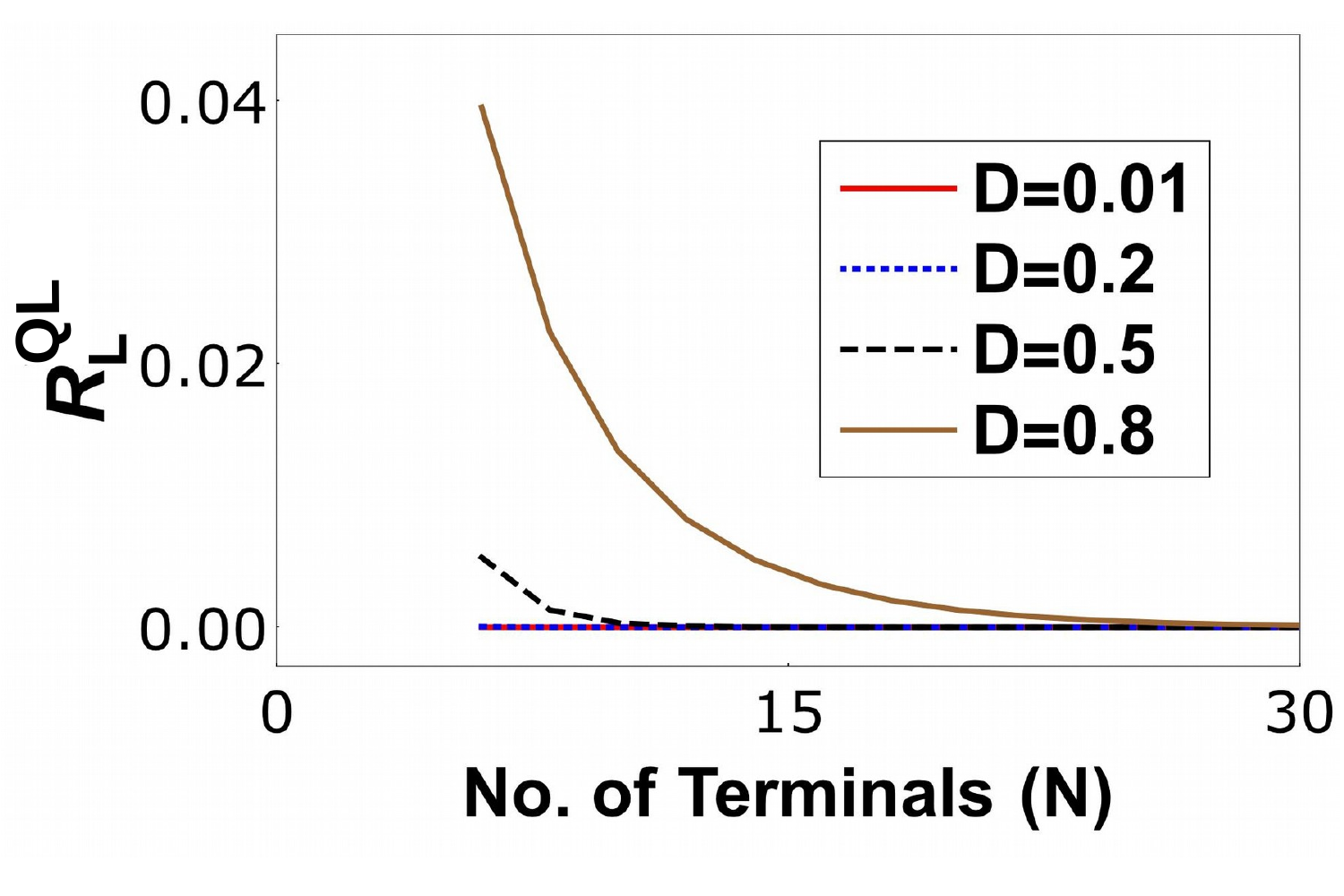}}
\vskip -0.2 in \caption{Longitudinal resistance in units of $\frac{h}{e^2}$  calculated (a) via scattering amplitudes, and (b) via probabilities for an N-terminal QSH sample with all contacts equally disordered, and (c) the quantum localization correction to the longitudinal resistance.}
\end{figure*} 
 
\begin{eqnarray}
R^{Amp}_H&=&\frac{h}{e^2}(D_u-D_l)*F,\nonumber\\
&=&0\quad(\text{when }D_l=D_u),\nonumber\\
R^{Amp}_{2T}&=&\frac{h}{e^2}\frac{(3-D(2-3D))(1+D^6-2D^3\cos\phi)}{2(1-D^2)^2(1+D^2+D^4)},\nonumber\\
R^{Amp}_L&=&\frac{h}{e^2}\frac{(1+D^6-2D^3\cos\phi)}{2(1+D)^2(1+D^2+D^4)},\nonumber\\
R^{Amp}_{NL}&=&\frac{h}{e^2}\frac{(1+D^6-2D^3\cos\phi)}{6(1+D)^2(1+D^2+D^4)},\nonumber
\end{eqnarray}
where,
\begin{equation}
F=\frac{(1 + D_l (2 - D_u) - 2 D_u) (1 + D_l^3 D_u^3 - 
    2 \sqrt{D_l^3 D_u^3} \cos\phi)}{
 6 (1 + D_l) (1 + D_u) (1 - D_l D_u)^2 (1 + D_l D_u (1 + D_l D_u))}.
\end{equation}
All contacts are considered to be equally disordered, i.e., $D_i=D$ (for $i=1-6$). To calculate the Hall resistance only we have considered $D_1=D_2=D_3=D_u$ and $D_4=D_5=D_6=D_l$, otherwise for equally disordered contacts the Hall resistance is always zero. After averaging over phase shift $\phi$ we get, 
\begin{eqnarray}
\langle R^{Amp}_H\rangle&=&\frac{h}{e^2}(D_u-D_l)*F',\nonumber\\&=&0,\quad(\text{when }D_l=D_u),\nonumber\\
\langle R^{Amp}_{2T}\rangle&=&\frac{h}{e^2}\frac{(3-D(2-3D))(1+D^6)}{2(1-D^2)^2(1+D^2+D^4)},\nonumber\\
\langle R^{Amp}_L\rangle&=&3\langle R^{Amp}_{NL}\rangle=\frac{h}{e^2}\frac{(1+D^6)}{2(1+D)^2(1+D^2+D^4)},\nonumber
\end{eqnarray}
where,
\begin{eqnarray}
F'=\frac{(1 + D_l (2 - D_u) - 2 D_u) (1 + D_l^3 D_u^3)}{
 6 (1 + D_l) (1 + D_u) (1 - D_l D_u)^2 (1 + D_l D_u (1 + D_l D_u))}.\nonumber\\
\end{eqnarray}
The quantum localization correction is the difference between the resistances calculated using probabilities, i.e., neglecting the phase acquired by the edge electrons and the resistance determined from scattering amplitudes, Eq.~(13). The conductance matrix $G$ derived from scattering probabilities is-
{\small
\begin{equation} 
 G=\frac{2e^2}{h}\frac{1}{b''}\left(\begin{smallmatrix}
2(1-R^5)T&-T^2R^4&-T^2R^3&-T^2R^2&-T^2R&-T^2\\
 -T^2&2(1-R^5)T&-T^2R^4&-T^2R^3&-T^2R^2&-T^2R\\
 -T^2R&-T^2&2(1-R^5)T&-T^2R^4&-T^2R^3&-T^2R^2\\
 -T^2R^2&-T^2R&-T^2&2(1-R^5)T&-T^2R^4&-T^2R^3\\
 --T^2R^3&-T^2R^2&-T^2R&-T^2&2(1-R^5)T&-T^2R^4\\
 -T^2R^4&-T^2R^3&-T^2R^2&-T^2R&-T^2&2(1-R^5)T\end{smallmatrix}\right),
 \end{equation}}
where $b''=(1-R^6)$. As before, current through voltage probes $2,3,5,6$ is zero, and choosing reference potential $V_4=0$ we get potentials $V_2$ and $V_4$ in terms of $V_1$. Thus, Hall resistance $R^{}_H$, 2-terminal resistance $R^{}_{2T}$, longitudinal resistance $R^{}_{L}$, and non-local resistance $R^{}_{NL}$ calculated via probabilities are then-
 \begin{eqnarray}
R^{}_H&=&\frac{(D_u- D_l) (1 + 2 D_u - D_l (2 + D_u))}{
 6 (1 + D_l) (1 + D_u) (1 - D_l D_u)},\nonumber\\&=&0, \text{ when }D_u=D_l,\nonumber\\
  R^{}_{2T}&=&\frac{h}{2e^2}\frac{(3-D(2-3D))}{2(1-D^2)},\nonumber\\ \text{and }R^{}_{L}&=&3R^{}_{NL}=\frac{h}{2e^2}\frac{(1-D)}{2(1+D)}.
\end{eqnarray}
The quantum localization corrections to the above calculated Hall, longitudinal, 2-terminal and non-local resistances in the 6-terminal QSH sample thus are $R_X^{QL}=\langle R^{Amp}_X\rangle-R_X^{}$, with $X=H, 2T, NL, L$-
\begin{eqnarray}
R^{QL}_H&=&\frac{(D_u - D_l) (1 - D_l D_u + 2 D_l^4 D_u^3 - 2 D_l^3 D_u^4)}{3 (1 + D_l) (1 + 
   D_u) (1 - D_l D_u)^2 (1 + D_l D_u + D_l^2 D_u^2)}\nonumber\\&=&0, \quad\text{when }D_u=D_l,\nonumber\\
R^{QL}_{2T}&=&\frac{h}{2e^2}\frac{D^6(3-2D+3D^2)}{(1-D^2)^2(1+D^2+D^4)},\nonumber\\
\text{and }R^{QL}_{L}&=&3R^{QL}_{NL}=\frac{h}{2e^2}\frac{D^6}{(1+D)^2(1+D^2+D^4)}.\nonumber\\
\end{eqnarray}
From Eq.~(17) we see that the quantum localization correction for Hall resistance in a six terminal QSH sample can be positive as well as negative depending on the strength of disorder at different contacts while for four terminal QSH sample it is always positive, see Eq.~(10). This negative correction term does not imply anti-localization of the helical electrons, rather it comes from the fact that the Hall resistance for QSH sample itself can be negative. However, the absolute value of resistances calculated via amplitudes is always greater than the absolute value of the resistances derived via probabilities, i.e., $|\langle R^{Amp}_H\rangle|>|R_H|$. This negative quantum localization correction for Hall resistance is unique to QSH samples only and not present for QH samples, see Ref.~\cite{arjun3}. From Eq.~(17) it can also be noted that for equally disordered contacts the quantum localization correction for Hall resistance vanishes for QSH samples while for QH samples it is finite, see Ref.~\cite{arjun3}. The quantum localization correction to the 2-terminal, longitudinal and non-local resistances increases with increasing disorder while the same for Hall resistance increases with the increase of the difference between the disorderedness of upper ($D_u$) and lower ($D_l$) contacts. 
 \begin{figure*}
  \centering \subfigure[]{ \includegraphics[width=0.33\textwidth]{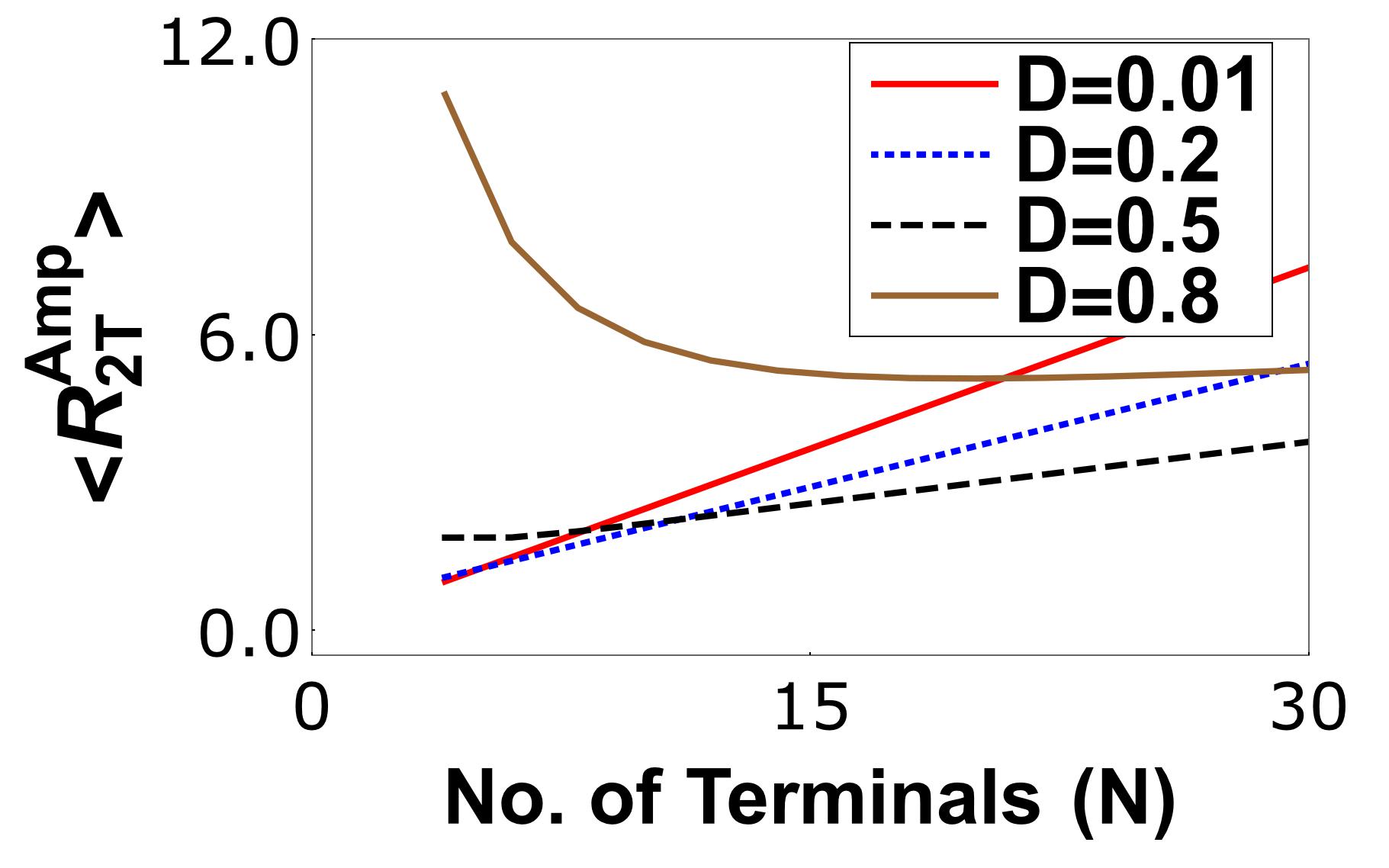}}
 \centering    \subfigure[]{ \includegraphics[width=0.32\textwidth]{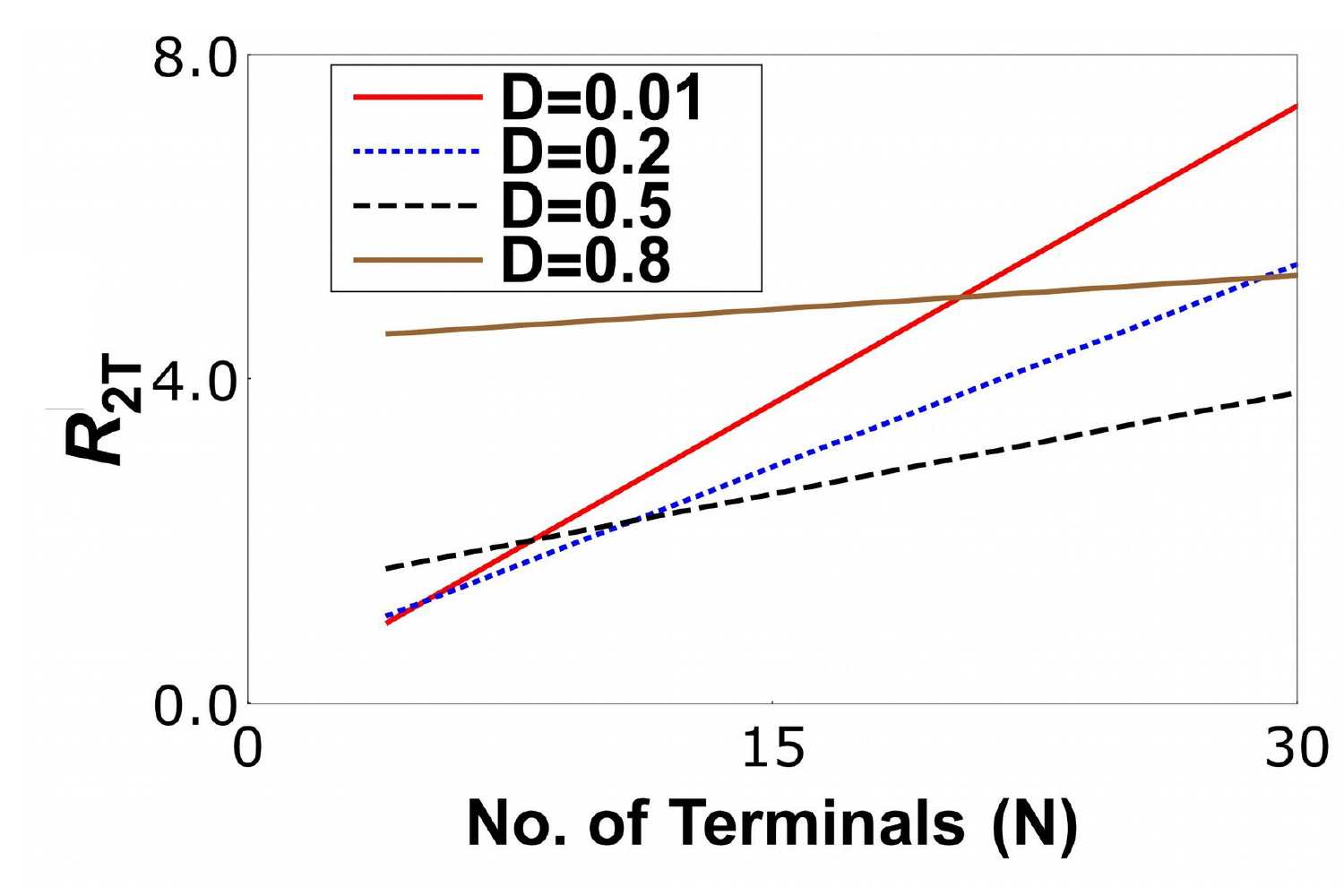}}
 \centering \subfigure[]{\includegraphics[width=.33 \textwidth]{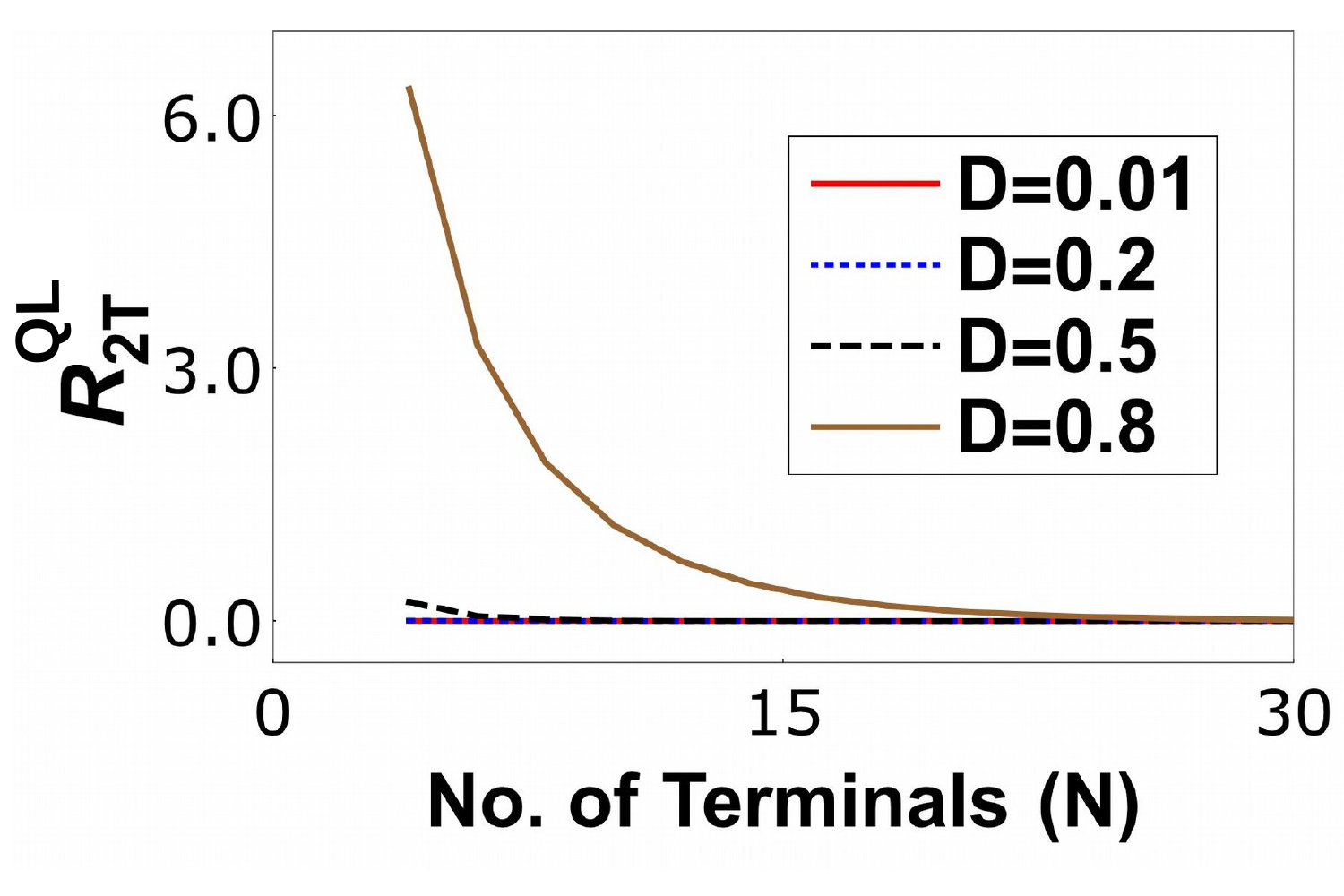}}
\vskip -0.2 in \caption{2-terminal resistance in units of $\frac{h}{e^2}$ calculated (a) via scattering amplitudes, (b) via probabilities  for a N-terminal QSH sample with all contacts equally disordered, and (c) the quantum localization correction to the 2T resistance.}
\end{figure*} 
  \begin{figure*}
  \centering \subfigure[]{ \includegraphics[width=0.33\textwidth]{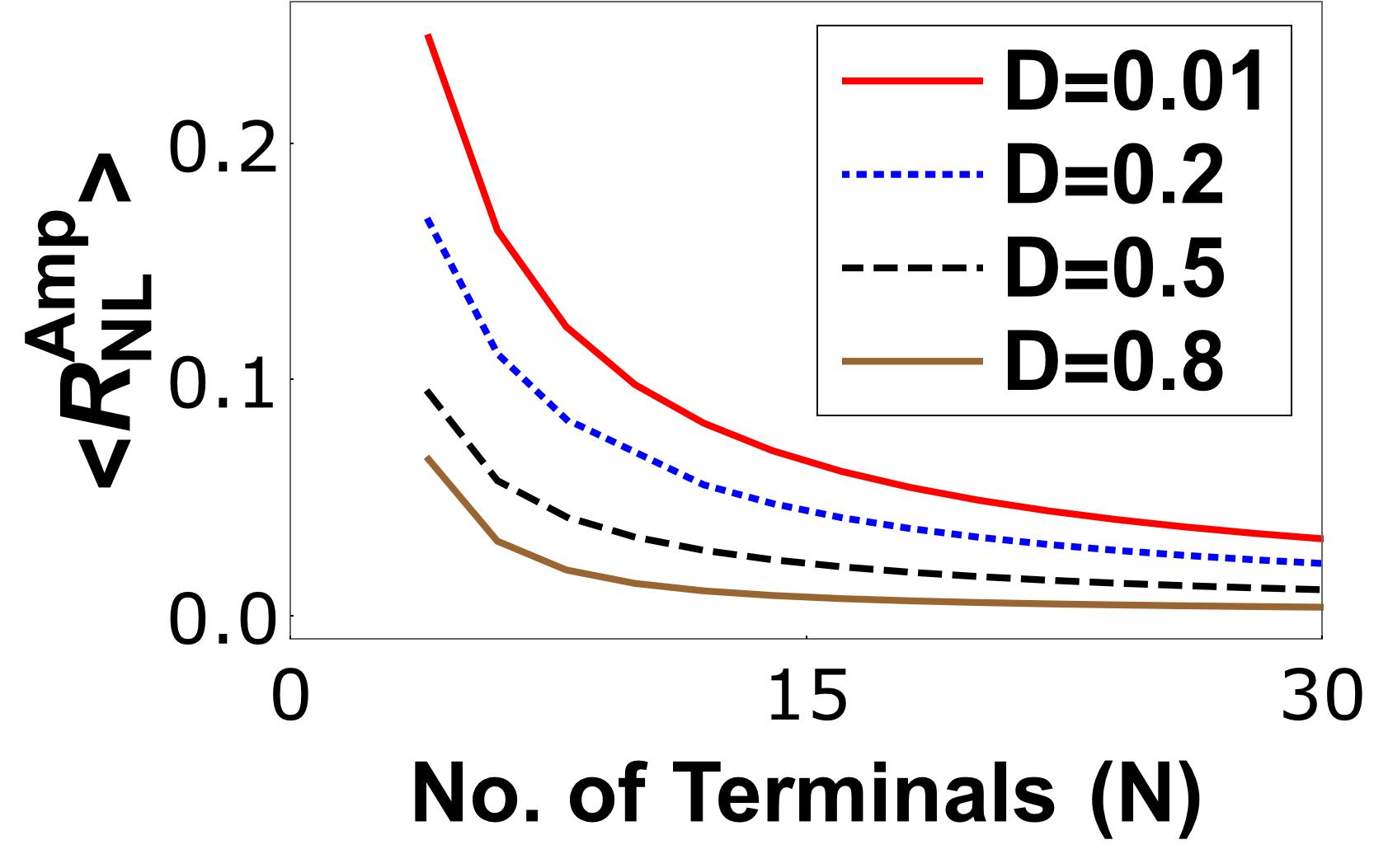}}
 \centering    \subfigure[]{ \includegraphics[width=0.32\textwidth]{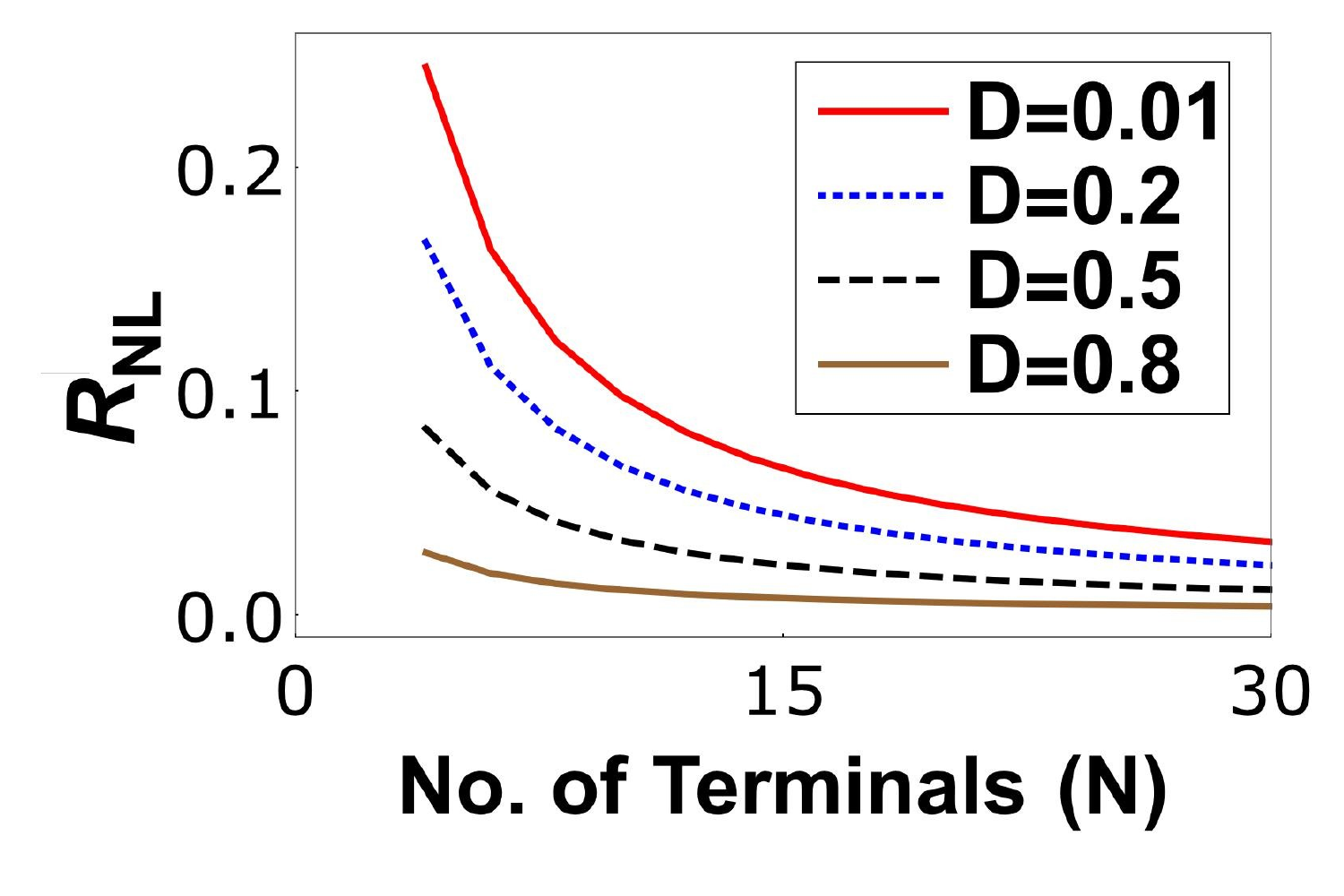}}
 \centering \subfigure[]{\includegraphics[width=.33 \textwidth]{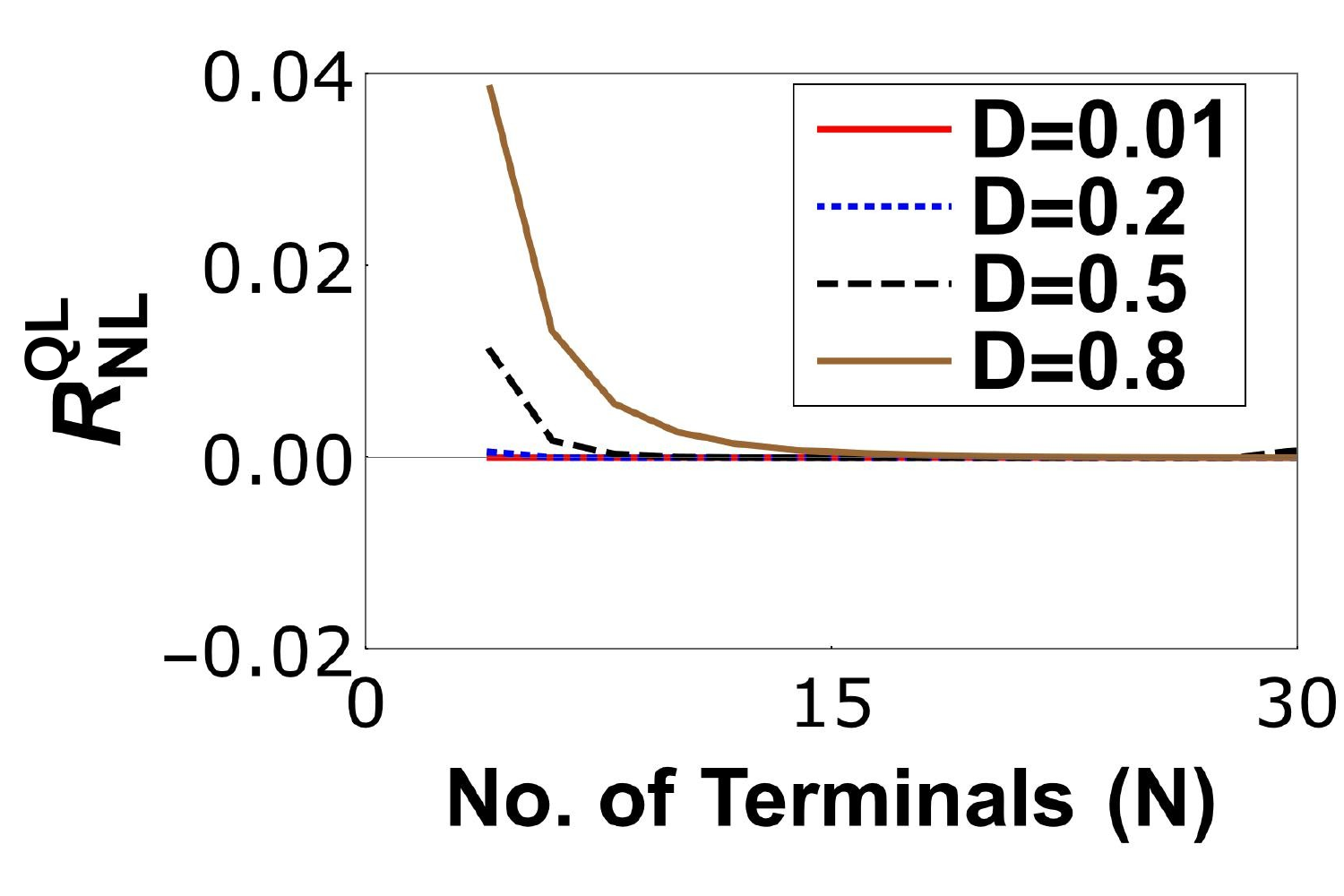}}
\vskip -0.2 in \caption{Non-local resistance in units of $\frac{h}{e^2}$ calculated (a) via scattering amplitudes, (b) via probabilities for a N-terminal QSH sample with all contacts equally disordered, and (c) the quantum localization correction to the non-local resistance.}
\end{figure*} 
 \section{N terminal system with all contacts disordered}
An N-terminal QSH sample is shown in Fig.~3(c) with all contacts equally disordered, i.e., $D_1=D_2=...=D_N=D$. Contacts $1$ and $k$ are current probes and contacts $2, 3,...k-1,k+1,...N$ are voltage probes, thus current through these contacts, i.e., $I_2=I_3=....=I_{k-1}=I_{k+1}=....=I_N=0$. The scattering matrix for the N-terminal QSH sample in Fig.~3(c) is
 \begin{widetext}
 {\small
\begin{equation} 
 S=\frac{1}{c}\left(\begin{smallmatrix}
 (r-r^{N-1}e^{\phi})e^{\phi_1}&0&...&-t^2r^{N-k}e^{i\phi_{k(k+1)..1}}&0&...&-t^2e^{i\phi_{N1}}&0\\
  0&(r-r^{N-1}e^{\phi})e^{\phi_1}&...&0&-t^2r^{k-2}e^{i\phi_{12..k}}&...&0&-t^2r^{N-2}e^{i\phi_{12..N}}\\
 .&.&...&.&.&...&.&.\\
  .&.&...&.&.&...&.&.\\
   -t^2r^{k-2}e^{i\phi_{12..k}}&0&...& (r-r^{N-1}e^{\phi})e^{\phi_k}&0&...&-t^2r^{k-1}e^{i\phi_{N12..k}}&0\\
   0&-t^2r^{N-k}e^{i\phi_{k(k+1)..1}}&...&0& (r-r^{N-1}e^{\phi})e^{\phi_k}&...&0&-t^2r^{N-k-1}e^{i\phi_{k(k+1)..N}}\\
      .&.&...&.&.&...&.&.\\
  .&.&...&.&.&...&.&.\\
   -t^2r^{N-2}e^{i\phi_{12..N}}& 0&...&-t^2r^{N-k-1}e^{i\phi_{k(k+1)..N}}&0&...&(r-r^{N-1}e^{\phi})e^{\phi_N}&0\\
   0& -t^2e^{i\phi_{N1}}&...&0&-t^2r^{k-1}e^{i\phi_{N12..k}}&...&0&(r-r^{N-1}e^{\phi})e^{\phi_N}
\end{smallmatrix}\right),
 \end{equation}
} \end{widetext}
where $c=1-r^Ne^{i\phi}$ and $\phi_{ij..k}=\phi_i+\phi_j+..+\phi_k$. The scattering matrix connects the incoming edge states to the outgoing edge states via the relation $(b_1^\uparrow,b_1^{\downarrow},...,b_k^\uparrow,b_k^\downarrow,...,b_N^\uparrow,b_N^\downarrow)^T=\mathcal{S} (a_1^\uparrow,a_1^{\downarrow},...,a_k^\uparrow,a_k^\downarrow,...,a_N^\uparrow,a_N^\downarrow)$. The conductance matrix $G$ of the N-terminal QSH sample derived from the scattering matrix $\mathcal{S}$, following Eq.~(1), is thus-
{
\begin{equation} 
 G=\frac{1}{c'}\left(\begin{smallmatrix}
 2T(1-R^{N-1})&...&-T^2(R^{N-k}+R^{k-2})&...&-T^2(1+R^{N-2})\\
 .&...&.&...&.\\
  .&...&.&...&.\\
   -T^2(R^{k-2}+R^{N-k})&...&2T(1-R^{N-1})&...&-T^2(R^{k-1}+R^{N-k-1})\\
      .&...&.&...&.\\
  .&...&.&...&.\\
   -T^2(R^{N-2}+1)&...&-T^2(R^{N-k-1}+R^{k-1})&...&2T(1-R^{N-1})
\end{smallmatrix}\right),
 \end{equation}
}
where $c'=1+R^{N}-2R^{N/2}\cos\phi$. Since currents through voltage probes $2,3,...,k-1,k+1,...,N$ is zero, so $I_2=I_3=...=I_{k-1}=I_{k+1}=I_N=0$, and choosing reference potential $V_k=0$ we get potentials $V_2$, $V_3$, $V_{k-1}$, $V_{k+1}$ and $V_N$ in terms of $V_1$. So, Hall resistance $R^{Amp}_H=R_{1k,2N}=\frac{(V_2-V_N)}{I_1}$, 2-terminal resistance $R^{Amp}_{2T}=R_{1k,1k}=\frac{(V_1-V_k)}{I_1}$, longitudinal resistance $R^{Amp}_L=R_{1k,23}=\frac{(V_2-V_3)}{I_1}$ and non-local resistance $R^{Amp}_{NL}=R_{12,(k+1)k}=\frac{(V_{k+1}-V_k)}{I_1}$. To calculate non-local resistance we consider contacts $1,2$ as current probes and contacts $3,4,..,k-1,k,k+1,...,N$ as voltage probe. As the expressions for these resistances are large, we analyze them via plots, see Figs.~(4-7). The average resistances for $N$-terminal case are found by averaging over the phases. Thus $\langle R^{Amp}_X\rangle=\frac{1}{2\pi}\int_{0}^{2\pi} R_Xd\phi$.  
 To calculate the quantum localization correction, we need to calculate the conductance using probabilities ignoring the phase acquired by the edge electrons. The conductance matrix $G$ derived via transmission probabilities is then
 {
\begin{equation} 
 G=\frac{1}{c''}\left(\begin{smallmatrix}
 2T(1-R^{N-1})&...&-T^2(R^{N-k}+R^{k-2})&...&-T^2(1+R^{N-2})\\
 .&...&.&...&.\\
  .&...&.&...&.\\
   -T^2(R^{k-2}+R^{N-k})&...&2T(1-R^{N-1})&...&-T^2(R^{k-1}+R^{N-k-1})\\
      .&...&.&...&.\\
  .&...&.&...&.\\
   -T^2(R^{N-2}+1)&...&-T^2(R^{N-k-1}+R^{k-1})&...&2T(1-R^{N-1})
\end{smallmatrix}\right),
 \end{equation}
}
where $c''=(1-R^{N})$. Setting the current, as before, through voltage probes $2,3,...,k-1,k+1,...,N$ to zero, and choosing reference potential $V_k=0$ we get potentials $V_2$, $V_3$, $V_{k-1}$, $V_{k+1}$ and $V_N$ in terms of $V_1$. Similarly, we need to calculate the Hall resistance $R^{}_H$, 2-terminal resistance $R^{}_{2T}$, and nonlocal resistance $R^{}_{NL}$ via probabilities from the conductance matrix as in Eq.~(20). As these expressions are large, we analyze them in Figs.~(4-7). The quantum localization correction, as defined before, is $R_X^{QL}=\langle R^{Amp}_X\rangle-R_X^{}$ with $X=H, 2T, L, NL$. One can get a closed form expression for a general $N$(with $N=$even)-terminal system as well by looking at the $ 6,  8, 10...$ terminal resistances. This is written below for the quantum localization correction, resistance derived via probabilities and that derived from amplitudes in case of longitudinal and non-local resistances-
\begin{eqnarray}
R^{Amp}_{L}&=&\frac{N}{2}R^{Amp}_{NL}=\frac{h}{2e^2}\frac{1+D^N-2D^{N/2} \cos\phi}{(1+D)^2 2(1+D^2+D^4+...+D^{N-2})},\nonumber\\
\langle R^{Amp}_{L}\rangle&=&\frac{N}{2}\langle R^{Amp}_{NL}\rangle=\frac{h}{2e^2}\frac{1+D^N}{(1+D)^2 2(1+D^2+D^4+...+D^{N-2})},\nonumber\\
R^{}_{L}&=&\frac{N}{2}R^{}_{NL}=\frac{h}{2e^2}\frac{(1-D)}{2(1+D)},\nonumber\\
R^{QL}_{L}&=&\frac{N}{2}R^{QL}_{NL}=\frac{h}{2e^2}\frac{D^N}{(1+D)^2(1+D^2+D^4+...+D^{N-2})}.
\end{eqnarray}
For simplicity, we consider in Eq.~(21) all contacts to be equally disordered. No closed form expression can be systematically deduced for Hall and 2-terminal cases as there is no uniformity in going from 6, 8, 10 terminal and likewise cases. In Figs.~(4-7) we analyze the quantum localization correction for various resistances. In Figs.~4(a,b), we see that Hall resistance for QSH case can be either negative or positive depending on disorder strength at upper edge ($D_u$) and lower edge ($D_l$) contacts. If $D_u=D_l$, then Hall resistance is zero in case of calculation using scattering amplitudes or probabilities. In Fig.~4(c) we see that the quantum localization correction to the Hall resistance can also be negative, which again does not imply that it leads to anti-localization. The Hall resistance for QSH itself can be negative, and that leads to a negative localization correction term, although $|\langle R_H^{Amp}\rangle|>|R^{}_H|$. The longitudinal resistance is almost constant as function of the number of contacts. However, the stronger the disorderedness of contacts the lower the longitudinal resistance. In Fig.~5(c), we see that the quantum localization correction decreases with increase in number of contacts unlike in quantum Hall samples where it is always zero, see Ref.~\cite{arjun3}. In Fig.~6(a,b) we see that the 2-terminal resistance for QSH case increases with number of contacts (unlike the QH case), which implies the 2-terminal resistance increases as a function of the length of the sample. This is similar to what is observed for Ohmic behavior. In Fig.~6(c) we see that the quantum localization correction is very small for $D<1/2$, only for $D>1/2$ it becomes substantial. In Figs.~4-7 we see that for large number of terminals the quantum localization correction disappears. Quantum localization correction is substantial only for strong disorder and few terminals.
\begin{figure}
  \centering{\includegraphics[width=0.45\textwidth]{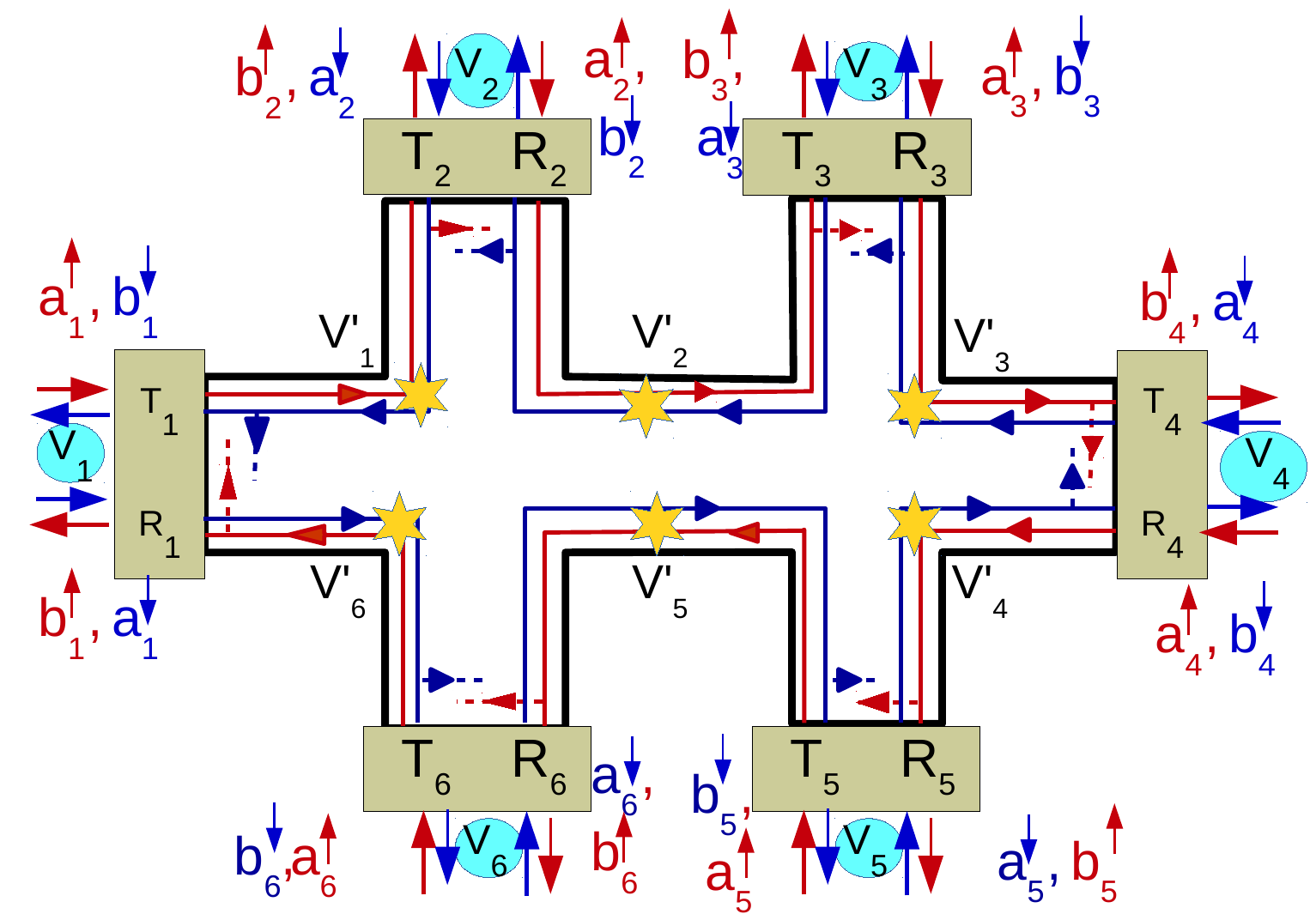}}
 \vspace{-.3cm}
\caption{6 terminal QSH bar with all disordered contacts and inelastic scattering.}
\end{figure}
\section{Effect of inelastic scattering on quantum localization correction}
A 6-terminal QSH sample with all disordered contacts and with inelastic scattering is shown in Fig.~8. When the length between the disordered contacts is larger than the phase coherence length for electronic edge modes, inelastic scattering occurs. In presence of inelastic scattering spin up edge electrons coming out of contact $1$ equilibrate with other spin up and down electrons at equilibrating potential $V_1'$ and lose their phase acquired via scattering at the contacts via equilibration of their energy. Similarly spin down electrons coming out of contact $1$ lose their phase at equilibrating potential $V_6'$ via equilibration of their energies with other spin up and down electrons. Thus, there is no possibility for an electron in a edge state to get back to the same contact after emerging out of it at that energy and with an unique phase. Thus, there is no difference between resistances calculated via probabilities and that via amplitudes. This implies absence of quantum localization correction in presence of inelastic scattering. Using probabilities the resistances have already been derived, see Refs.~\cite{arjun1,arjun2}, as- 
\begin{eqnarray}
R_H&=&0,\quad
R_{2T}=\frac{h}{e^2}\frac{(3-D)}{(1-D^2)}\quad
R_L=3R_{NL}=\frac{h}{e^2}\frac{1}{(1+D)},\nonumber\\
\end{eqnarray}
with $R_X=\langle R^{Amp}_X\rangle$, $X=H,L,2T, NL$. Here, we have only concentrated on the six terminal QSH system, as in 4- and N-terminal QSH sample we obtain exactly similar results wherein inelastic scattering completely kills the quantum localization correction. 

{
\begin{center}
{ Table1: Comparison of the quantum localization correction in 6-terminal QH \cite{arjun3} and QSH sample with equally disordered contacts} \\
\begin{tabular}{ |c|c|c|} 
 \hline
& {Quantum Hall}  &{Quantum spin Hall}  \\ 
\hline 
 \hline
$R_H^{QL}$&$\frac{h}{2e^2}\frac{2D^6}{1-D^6}$&0\\
\hline
$R_L^{QL}$&0&$\frac{h}{2e^2}\frac{D^6}{(1+D)^2(1+D^2+D^4)}$\\
\hline
$R_{2T}^{QL}$&$\frac{h}{2e^2}\frac{2D^6(1+D)}{(1-D)(1-D^6)}$&$\frac{h}{2e^2}\frac{D^6(3-2D+3D^2)}{(1-D^2)^2(1+D^2+D^4)}$\\
\hline
$R_{NL}^{QL}$&0&$\frac{h}{2e^2}\frac{D^6}{3(1+D)^2(1+D^2+D^4)}$\\
\hline
\hline
\end{tabular}
\end{center}
}
\section{Conclusion}
We see that resistances are affected by the quantum localization correction but only when all contacts are disordered. The quantum localization correction for the resistances  for both QH (see Ref.~\cite{arjun3}) and QSH six terminal samples are summarized and compared in Table~1. From Table~1, we see that for equally disordered contacts in QH sample only 2-terminal and Hall resistances are affected by the quantum localization correction, while in QSH sample the 2-terminal, longitudinal and non-local resistances are affected by the same correction. { Quantum localization correction term arises in a QSH or QH sample due to multiple paths available edge mode electrons due to the fact that all contacts are disordered as explained in section II. However, summing the multiple paths available for helical edge modes in QSH samples and chiral edge modes in QH sample leads to a difference in the quantum localization correction. A remark on the table- the vanishing quantum localization correction doesn't mean $\langle R^{Amp}_{L,NL}\rangle = R^{}_{L,NL}$ for a QH sample or $\langle R^{Amp}_{H}\rangle= R^{}_{H}$ for a QSH sample but rather because $\langle R^{Amp}_{L,NL}\rangle= R^{}_{L,NL}=0$ for QH sample and same for Hall resistance in QSH sample. This suggests that the quantum localization correction term is finite only when resistances calculated via scattering amplitudes or probabilities are themselves finite. }

In QSH samples we even see a negative localization correction, which is not due to the anti localization of the states, but rather due to the fact that the Hall resistance in a QSH system can itself turn negative. In presence of inelastic scattering this quantum localization term  vanishes for both QH and QSH cases. {In this letter, we have assumed disorder only at the contacts, there is no disorder within the sample. Generally, edge modes in QH/QSH samples suffer some amount of scattering at contacts. The presence of disorder within the sample wont affect the results of our letter, since it is well known that QH and QSH edge modes are robust to sample disorder. Disorder at contacts works as a barrier to edge mode transport, edge modes can partially transmit into the contacts through the barrier with probability $T$ or can be partially reflected with probability $R$. In case one has completely clean contacts, one can design sample contacts to  partially reflect edge modes at contacts by directly doping non-magnetic impurities or via creating an electrostatic barrier at the contacts. In Refs.~\cite{john, wanli}, the authors have studied sample disorder in quantum Hall systems via doping impurities within the sample. Similarly, impurities can be doped into contacts in a QSH sample thus realizing our setups and verifying the quantum localization correction.} 

\acknowledgments
This work was supported by funds from SERB, Dept. of Science and Technology, Government of India, Grant No. EMR/2015/001836.
  
\end{document}